\documentclass[12pt]{elsarticle}

\usepackage{soul} 

\usepackage{placeins}

\usepackage{amsmath,amsfonts,amssymb,mathrsfs,tensor,braket,dsfont}
\usepackage{dirtree}
\usepackage{listings}

\renewcommand{\vec}[1]{\boldsymbol{#1}}

\newcommand{\cu}{\mathbf{i}}
\newcommand{\e}{\mathrm{e}}
\newcommand{\dif}[1][d]{\, \mathrm{#1}} 
\newcommand{\comm}[1]{}
\newcommand{\E}{E}

\newcommand{\p}{\partial}
\newcommand{\pd}[2]{\frac{\partial {#1}}{\partial {#2}}}

\let\Re\undefined
\let\Im\undefined

\DeclareMathOperator{\Im}{\mathsf{Im}}
\DeclareMathOperator{\Re}{\mathsf{Re}}

\DeclareMathOperator{\Arg}{\mathrm{Arg}}

\usepackage{xcolor}

\journal{CPC}

\usepackage[edges]{forest}

\colorlet{folderbg}{orange}
\colorlet{folderborder}{black}

\def\Size{4pt}
\tikzset{
  folder/.pic={
    \filldraw[draw=folderborder,top color=folderbg!50,bottom color=folderbg]
      (-1.05*\Size,0.2\Size+5pt) rectangle ++(.75*\Size,-0.2\Size-5pt);  
    \filldraw[draw=folderborder,top color=folderbg!50,bottom color=folderbg]
      (-1.15*\Size,-\Size) rectangle (1.15*\Size,\Size);
  },
  file/.pic={%
    \filldraw [draw=folderborder, top color=folderbg!5, bottom color=folderbg!10] (-\Size,.4*\Size+5pt) coordinate (a) |- (\Size,-1.2*\Size) coordinate (b) -- ++(0,1.6*\Size) coordinate (c) -- ++(-5pt,5pt) coordinate (d) -- cycle (d) |- (c) ;
  },
}

\forestset{%
  declare autowrapped toks={pic me}{},
  declare boolean register={pic root},
  pic root=0,
  pic dir tree/.style={%
    for tree={%
      folder,
      font=\ttfamily,
      grow'=0,
    },
    before typesetting nodes={%
      for tree={%
        edge label+/.option={pic me},
      },
      if pic root={
        tikz+={
          \pic at ([xshift=\Size].west) {folder};
        },
        align={l}
      }{},
    },
  },
  pic me set/.code n args=2{%
    \forestset{%
      #1/.style={%
        inner xsep=2\Size,
        pic me={pic {#2}},
      }
    }
  },
  pic me set={directory}{folder},
  pic me set={file}{file},
}

\begin{document}

\begin{frontmatter}



\title{Modular multiscale approach to modelling high-harmonic generation in gases}


\affiliation[inst1]{organization={ELI Beamlines Facility, The Extreme Light Infrastructure ERIC},
            addressline={Za Radnicí 835}, 
            city={Dolní Břežany},
            postcode={25241}, 
            state={Czech Republic}}
\affiliation[inst2]{organization={Czech Technical University in Prague, Faculty of Nuclear Sciences and Physical Engineering},
            addressline={Břehová 7}, 
            city={Prague 1},
            postcode={11519}, 
            state={Czech Republic}}
\affiliation[inst3]{organization={Université Claude Bernard Lyon 1, CNRS, Institut Lumière Matière, UMR5306},
            city={Villeurbanne},
            postcode={F-69100}, 
            state={France}}
\affiliation[inst4]{organization={Centre Lasers Intenses et Applications, Université de Bordeaux-CNRS-CEA},
            city={Talence Cedex},
            postcode={33405}, 
            state={France}}

\author[inst1,inst2]{Jan Vábek\corref{cor1}}
\ead{jan.vabek@eli-beams.eu}
\ead{vabekjan@fjfi.cvut.cz}
\author[inst1,inst2]{Tadeáš Němec}
\author[inst3]{Stefan Skupin}
\ead{stefan.skupin@cnrs.fr}
\author[inst4]{Fabrice Catoire}
\ead{fabrice.catoire@u-bordeaux.fr}

\cortext[cor1]{Corresponding author}

\begin{abstract}
We present a~modular user-oriented simulation toolbox for studying high-harmonic generation in gases~\cite{Vabek_Multiscale_modular_approach_2025}. The first release consists of the computational pipeline to 1)~compute the unidirectional IR-pulse propagation in cylindrical symmetry, 2)~solve the microscopic responses in the whole macroscopic volume using a~1D-TDSE solver, 3)~obtain the far-field harmonic field using a~diffraction-integral approach. The code comes with interfaces and tutorials, based on practical laboratory conditions, to facilitate the usage and deployment of the code both locally and in HPC-clusters. Additionally, the modules are designed to work as stand-alone applications as well, e.g., 1D-TDSE is available through Pythonic interface.
\end{abstract}

\end{frontmatter}


\section{Introduction}

This computational toolbox results from a~joint effort between the high harmonic generation~(HHG) and \comm{ultrashort-filament} fs-laser filamentation communities. It combines 1)~a~code initially designed to study the reshaping and filamentation due to non-linear laser pulse propagation, and 2)~codes for solving the time-dependent Schrödinger equation~(TDSE) and for modelling HHG and simulation of the subsequent linear XUV-field propagation. 
The result is a~modular codebase that facilitates the modelling of HHG experiments, encompassing both the macroscopic and microscopic aspects of the phenomenon. The first release~\cite{Vabek2026_zenodo} is limited to modelling under linear polarisation and cylindrical symmetry during propagation. The primary goal is to establish a detailed~one-to-one correspondence with laboratory HHG experiments. However, the modules are also designed to be used independently to address different aspects of HHG and related areas of atomic and plasma physics. Before delving into the specifics of the code, let us first explore the physical context of HHG.

HHG in a~gaseous medium produces coherent XUV radiation and is the result of a highly non-linear response of the medium induced by an infrared~(IR) driving laser pulse~\cite{Lewenstein1994, Calegari2016, Weissenbilder2022, Castelvecchi2023-jr, Constant2025}. 
The fundamental physical mechanism behind HHG is the microscopic interaction of a~single atom with the external laser field, explained by the three-step model~\cite{Corkum1993} that is rigorously encompassed by solving the TDSE. However, the efficiency of IR-to-XUV conversion remains low, which constitutes a~challenge for applications requiring high-intensity XUV pulses, such as XUV–XUV pump–probe experiments~\cite{Jiang2010} or non-linear ionization~\cite{Benis2006}. 
However, the realisation of the process usually occurs on a~macroscopic scale corresponding to the laboratory environment using a~gas cell or jet as the generating medium. This brings additional complexity due to the non-linear IR dynamics during the propagation such as plasma defocusing that reshapes the driving pulse. This dynamics then naturally impacts the HHG at different macroscopic points in the medium. The description of XUV propagation is less complicated due to its low intensity, as significant self-induced non-linear reshaping is ruled out. In summary, the coupling of microscopic and macroscopic aspects remains computationally challenging and of particular interest for applications.

\begin{figure}[ht]
\centering
\includegraphics[width=\textwidth]{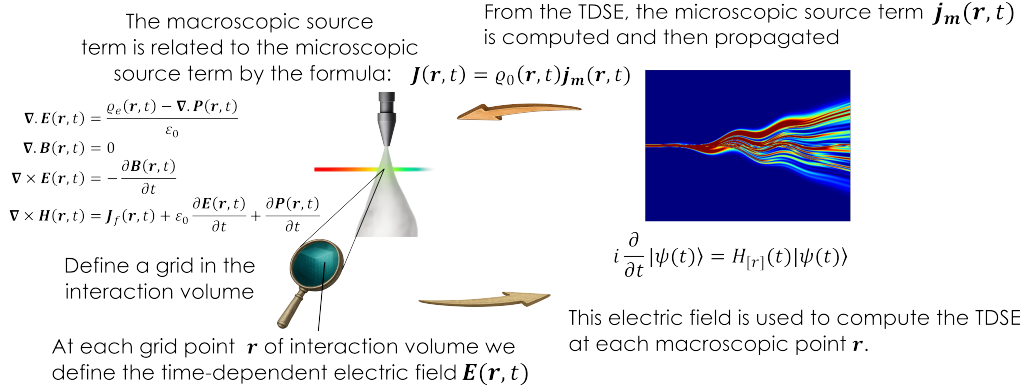}
\caption{General MMA scheme showing the coupling of the macroscopic field propagation governed by the macroscopic Maxwell's equations together with microscopic dynamics given by TDSE. It presents the idea of the full feedback loop, which poses a~great computational challenge in a~general case.}
\label{fig:MMA_scheme_intro}
\end{figure}

Generally speaking, the non-linear propagation of a~pulse in a~medium would require a~feedback loop coupling the evolutions of the electromagnetic field and quantum-mechanical treatment of the response of the medium~\cite{Lorin2007}. This loop is illustrated in Fig.~\ref{fig:MMA_scheme_intro}. This idea is similar to the key concept introduced for particle-in-cell codes (see, e.g.,~\cite{Derouillat2018}), except that the kinetic treatment of charged particles is replaced by the microscopic response of the electrons within the influence of atomic nuclei. However, this solution constitutes a~significant computational challenge which could be avoided without too constraining assumptions, as we will present in this manuscript.

The present model originates from the integration of multiple numerical codes developed within the high-harmonic generation (HHG) and femtosecond laser filamentation communities. The IR propagation was first described by a scalar envelope model in cylindrical symmetry~\cite{Skupin2002,Nuter2005,Berg2005,Skupin2006,Skupin:oc:280:173,Berge:prl:101:213901} or fully space- and time resolved~\cite{Berg2004,Skupin2004a,Skupin2004b,Sohr2021}, which is sufficient for many applications involving not too short laser pulses. The code was then upgraded to the unidirectional pulse propagation equation (UPPE)~\cite{Kolesik2004} to be able to describe larger frequency bandwidths~\cite{Babushkin2010,Babushkin2011,Berg2013,Martnez2015}, and a vectorial model was also implemented~\cite{Tailliez2020,Stathopulos2021,Stathopulos2023}. Early HHG approaches were mainly restricted to regimes in which macroscopic propagation effects and reshaping of the driving-field could be safely neglected~\cite{Catoire2016, Quintard2019, Veyrinas2021}. Subsequent developments progressively relaxed these assumptions, leading to models that self-consistently account for the propagation-induced evolution of the driving field~\cite{Finke2022}. The direction for future development is to use the code to provide more detailed insight into advanced beam geometries~\cite{Finke2024} or to extend the capability to study generating schemes requiring vectorial fields~\cite{Picot2025}.

The toolbox addresses the different HHG subproblems through a collection of dedicated codes, each focussing on a specific component:

\begin{enumerate}
	\item \textbf{The non-linear laser propagation}: This is computed under cylindrical symmetry using the CUPRAD (Complex Unidirectional Propagator, RADial symmetry in the transverse plane) code. This is the first module in the chain and can be run independently.    
	\item \textbf{The microscopic response}: The microscopic response is obtained by numerically solving the TDSE. This code is compiled as an MPI application and a dynamic library which can be accessed directly from Python.    
	\item \textbf{XUV propagation}: The XUV field propagation is modelled using a diffraction integral. This module is implemented in Python and can be fed with any user-defined microscopic dipole.   
\end{enumerate}

These modules can be chained together, forming a pipeline that offers a comprehensive multiscale description of HHG (see Fig.~\ref{fig:MMA_scheme2}). Data exchange between modules is facilitated by the HDF5 format, ensuring that all inputs and outputs are consistent in one file and easy to manipulate. Thus, the whole simulation pipeline can be executed at once.

This article serves as the top layer of the documentation, discussing the physical context and computational methods used in the code. It covers:

\begin{itemize}
\item IR propagation (Section~\ref{Infrared Propagation});	\item High Harmonic Generation (HHG) and propagation (Section~\ref{High Harmonic Generation and Propagation}); 
\item Details of the implementation of the MMA scheme (Section~\ref{A Modular Multiscale Approach (MMA)}); 
\item Details of the user interface (Section~\ref{User Interfaces}); \item Usage of the Pythonic CTDSE library (Section~\ref{Interactive Python Interface to CTDSE}); 
\item Practical examples to demonstrate usage (Section~\ref{Physics Examples}).
\end{itemize}
In addition to this paper, the complete documentation of the modules is available in the associated \emph{Git repository}~\cite{Vabek_Multiscale_modular_approach_2025, Vabek2026_zenodo}. This repository includes Jupyter notebooks that show the use of the code (see also Section~\ref{Physics Examples}). The key parts of the model are documented directly in the source code of the computational routines. Additionally, the repository provides guidelines for using the code within Docker containers and on HPC clusters.

For the sake of clarity, we use the symbol~$\hat{f}$ to represent the temporal Fourier transform of~$f$. All physical descriptions are given in SI units, except for the microscopic physics sections where different units are used (Sections~\ref{sec:High Harmonic Generation and Propagation:Microscopic Response in XUV regime},~\ref{Implementations: CTDSE} and~\ref{Interactive Python Interface to CTDSE}).

\section{Infrared Propagation}
\label{Infrared Propagation}

In this section, we present the model equations used to describe the propagation of ultrashort ionising pulses in gases. For the parameter ranges considered here, a~\emph{complex envelope} formulation of the scalar optical field, based on an extended nonlinear Schr\"odinger equation in cylindrical coordinates, provides an adequate description \cite{Bergé2007,COUAIRON200747}.
The propagation is expressed in terms of the complex field envelope using the slowly varying envelope approximation (SVEA) in both space and time, while allowing for higher-order effects such as space–time coupling and self-steepening \cite{Brabec_prl_78_3282}.
Future versions of the code will incorporate the unidirectional pulse propagation equation \cite{Kolesik2004}, which does not rely on the SVEA and enables the modelling of few-cycle pulses. The current restriction to linearly polarised (scalar) fields and cylindrical symmetry will also be relaxed in future releases.
The following section provides a brief outline of the propagation models implemented in the current version of CUPRAD.

Our starting point is macroscopic Maxwell's equations
\begin{subequations}
\label{maxwell}
\begin{align}
\nabla \cdot \vec{E}(\vec{r},t) & = \frac{\varrho(\vec{r},t) - \nabla \cdot \vec{P}(\vec{r},t)}{\epsilon_0} \,, \label{Maxwell1}\\
\nabla \cdot \vec{B}(\vec{r},t) & = 0 \label{Maxwell2} \,,\\
\nabla \times \vec{E}(\vec{r},t) & = - \frac{\partial}{\partial t} \vec{B}(\vec{r},t)\,, \label{Maxwell3}\\
\nabla \times \vec{H}(\vec{r},t) & = \vec{J}_f(\vec{r},t) + \epsilon_0 \frac{\partial}{\partial t} \vec{E}(\vec{r},t) + \frac{\partial}{\partial t} \vec{P}(\vec{r},t)\,. \label{Maxwell4}
\end{align}
\end{subequations}
All fields involved—namely the electric field $\vec{E}$, the polarisation vector $\vec{P}$, the magnetic field $\vec{H}$, the magnetic induction vector $\vec{B}$, the macroscopic charge density $\varrho$ and the corresponding current density $\vec{J}_f$—are real-valued quantities. Each of these fields is physically associated with specific processes: The polarisation $\vec{P}$ represents the microscopic bound response of the medium, encompassing both bound–bound contributions related to nonlinear $\chi^{(n)}$ processes and bound–continuum interactions such as high-harmonic generation (HHG). The current density $\vec{J}_f$, on the other hand, accounts for the dynamics of free carriers, particularly those generated by ionisation. A more detailed description of these quantities is provided in Section~\ref{sec:High Harmonic Generation and Propagation:Microscopic Response in XUV regime}.

Gases are optically isotropic and non-magnetisable media that exhibit a nonlinear polarisation vector. Provided that the spectral range of the laser field lies sufficiently far from any material resonances, the standard framework of nonlinear optics can be applied. In this regime, the polarisation vector $\vec{P}$ can be expressed as a power series in $\vec{E}$:
\begin{subequations}
\begin{align}
\widehat{\vec{P}}(\vec{r},\omega) & = \widehat{\vec{P}}\raisebox{1ex}{\scriptsize(1)}(\vec{r},\omega) + \widehat{\vec{P}}\raisebox{1ex}{\scriptsize(3)}(\vec{r},\omega) + \widehat{\vec{P}}\raisebox{1ex}{\scriptsize(5)}(\vec{r},\omega) + \widehat{\vec{P}}\raisebox{1ex}{\scriptsize(7)}(\vec{r},\omega) + \ldots \label{polarization}\\
\widehat{P}_{\mu}^{(j)}(\vec{r},\omega) & = \epsilon_0 \sum_{\alpha_1 \ldots \alpha_j} \idotsint \chi_{\mu \alpha_1 \ldots \alpha_j}^{(j)}(\vec{r},-\omega_{\sigma};\omega_1,\ldots,\omega_j) \label{chij}\\ 
& \quad\times \widehat{E}_{\alpha_1}(\vec{r},\omega_1) \ldots \widehat{E}_{\alpha_j}(\vec{r},\omega_j) \delta(\omega-\omega_{\sigma}) d\omega_1 \ldots d\omega_j \nonumber \\
\omega_{\sigma} & = \omega_1 + \ldots + \omega_j.
\end{align}
\end{subequations}
For magnetic induction, we have
\begin{equation}
\widehat{\vec{B}}(\vec{r},\omega) = \mu_{0}\widehat{\vec{H}}(\vec{r},\omega). \label{nomag}
\end{equation}
For technical convenience, we express these relations in Fourier space and the corresponding transformations are defined as
\begin{align}
\widehat{\vec{F}}(\vec{r},\omega) & = \frac{1}{\sqrt{2\pi}}\int\vec{F}(\vec{r},t)e^{\cu\omega t}dt\\
\vec{F}(\vec{r},t) & = \frac{1}{\sqrt{2\pi}}\int\widehat{\vec{F}}(\vec{r},\omega)e^{-\cu\omega t}d\omega.
\end{align}
In Equation (\ref{polarization}), we account for the fact that, in isotropic media, all even-order susceptibility tensors $\overset{\leftrightarrow}{\chi}\raisebox{1ex}{\scriptsize(j)}$ with even $j$ vanish due to considerations of spatial symmetry \cite{Boyd:NLO}. The subscripts $\mu$, $\alpha_1$, $\ldots$, $\alpha_j$ in Equation (\ref{chij}) denote the respective field vector component in Cartesian coordinates. As indicated by the summation symbol, the indices $\alpha_1$, $\ldots$, $\alpha_j$ are summed over the three spatial directions $x$, $y$, and $z$.
In the following, we specify Equation (\ref{polarization}) for the media considered in this work. As will be shown later, for the field intensities relevant to our study, the condition $|\vec{P}^{(3)}| \ll |\vec{P}^{(1)}|$ is already satisfied. Consequently, we truncate the expansion (\ref{polarization}) after the third-order term.\footnote{CUPRAD also allows the inclusion of fifth-order nonlinearities; however, for clarity, they are omitted from the present derivation.}

The expression for linear polarisation $\widehat{\vec{P}}\raisebox{1ex}{\scriptsize(1)}$ can be further simplified. In an isotropic medium, the susceptibility tensor $\overset{\leftrightarrow}{\chi}\raisebox{1ex}{\scriptsize(1)}$ is diagonal, leaving only one independent element component: $\displaystyle{\chi_{\mu \alpha}^{(1)} = \chi_{xx}^{(1)} \delta_{\mu \alpha}}$. Hence, according to the convention $\displaystyle{\chi^{(1)}(\vec{r},\omega)=\chi_{xx}^{(1)}(\vec{r},-\omega;\omega)}$, we have
\begin{equation}
\widehat{\vec{P}}\raisebox{1ex}{\scriptsize(1)}(\vec{r},\omega) = \epsilon_0 \chi^{(1)}(\vec{r},\omega) \widehat{\vec{E}}(\vec{r},\omega), \label{linpol}
\end{equation}
and we define the scalar dielectric function
\begin{equation}
\epsilon(\vec{r},\omega) = 1 + \chi^{(1)}(\vec{r},\omega). \label{epsilon}
\end{equation}
The third order nonlinear polarisation vector $\widehat{\vec{P}}\raisebox{1ex}{\scriptsize(3)}$ is determined by the 81 components of the general fourth-rank tensor $\overset{\leftrightarrow}{\chi}\raisebox{1ex}{\scriptsize(3)}$. Once again, taking advantage of the isotropy of the medium and applying spatial symmetry relations \cite{Boyd:NLO}, it can be shown that
\begin{equation}
\chi_{\mu \alpha_1 \alpha_2 \alpha_3}^{(3)} = \chi_{xxyy}^{(3)}\delta_{\mu \alpha_1}\delta_{\alpha_2 \alpha_3} + \chi_{xyyx}^{(3)}\delta_{\mu \alpha_3}\delta_{\alpha_1 \alpha_2} + \chi_{xyxy}^{(3)}\delta_{\mu \alpha_2}\delta_{\alpha_1 \alpha_3}.
\end{equation}
Furthermore, we restrict our analysis to media that exhibit homogeneous nonlinearity and to linearly polarised electric fields of the form $\vec{E} = E_x \vec{e}_x$.
Under these conditions, only a single relevant component of the susceptibility tensor contributes, and with the definition
\begin{equation}
\chi^{(3)}=\chi_{xxxx}^{(3)}=\chi_{xxyy}^{(3)}+\chi_{xyyx}^{(3)}+\chi_{xyxy}^{(3)}
\end{equation}
we have
\begin{align}
\widehat{\vec{P}}\raisebox{1ex}{\scriptsize(3)}(\vec{r},\omega) & = \vec{e}_x \epsilon_0 \iint \chi^{(3)}(\omega;\omega_1,\omega_2,\omega-\omega_1-\omega_2) \label{chi3} \\
& \quad \times \widehat{E}_x(\vec{r},\omega_1)\widehat{E}_x(\vec{r},\omega_2)\widehat{E}_x(\vec{r},\omega-\omega_1-\omega_2)d\omega_1d\omega_2. \nonumber
\end{align}
As expected, in an isotropic medium the nonlinear polarisation vector is parallel to the electric field.

In media of interest, such as air or noble gases, the dielectric function $\epsilon(\omega)$ is real and positive for wavelengths extending  from infrared to ultraviolet. The linear refractive index is defined by $n(\omega)=\sqrt{\epsilon(\omega)}$, and the corresponding wavenumber is defined by $k(\omega)= \omega n(\omega) /c$. We consider the propagation of a pulsed beam with a spectrum centred around the operating frequency $\omega_0$. Assuming that the spectral bandwidth $\Delta \omega$ is sufficiently narrow, the wavenumber $k(\omega)$ can be expressed as a Taylor series expansion around $\omega _{0}$:
\begin{subequations}
\begin{align}
k(\omega) & =  k_{0}+k^{\prime }\overline{\omega }+\frac{1}{2}k^{\prime \prime }\overline{\omega }^{2}+\frac{1}{6}k^{\prime \prime \prime}\overline{\omega }^{3}+\ldots \,, \\
k^{2}(\omega) & = k_{0}^{2}+2k_{0}k^{\prime }\overline{\omega }+k^{\prime 2}\overline{\omega }^{2}+k_{0}k^{\prime \prime }\overline{\omega }^{2}+k^{\prime}k^{\prime \prime }\overline{\omega }^{3}+\frac{k_{0}k^{\prime \prime \prime}}{3}\overline{\omega }^{3}+\ldots \,. \label{nottingham_eq: ksq}
\end{align}
\end{subequations}
where $\overline{\omega }=\omega -\omega _{0}$. The complex slowly varying envelope function $\vec{\mathcal{E}}$ of the electric field $\vec{E}$ is defined by
\begin{equation}
\vec{E}(\vec{r},t) = \sqrt{\frac{\omega_{0}\mu_{0}}{2k_{0}}} \vec{\mathcal{E}} (x,y,z,t) e^{\cu(k_{0}z-\omega_0t)} + \textrm{c.c.}. \label{sveart}
\end{equation}
Note that this introduces the variable $\vec{\mathcal{E}}$ and no approximation is made at this stage; especially, the phase of~$\vec{\mathcal{E}}$ carries chirps and shifts of the fundamental frequency.
As already mentioned, we consider linearly polarised light, that is,
the $x$-component of the optical field is assumed to be the dominant one, $| \mathcal{E}_x | \gg | \mathcal{E}_{y,z} |$, and we write $\mathcal{E} = \mathcal{E}_x$. The field envelope $\mathcal{E}$ is assumed to change slowly in $\alpha=x$, $y$, $z$ and $t$
\begin{subequations}
\label{svea_spacetime}
\begin{eqnarray}
\left|\frac{\partial }{\partial \alpha}\mathcal{E}\right| & \ll &  k_{0}\left|\mathcal{E}\right| \label{sveaspace}\\
\left|\frac{\partial }{\partial t}\mathcal{E}\right| & \ll &  \omega_{0}\left|\mathcal{E}\right|. \label{sveatime}
\end{eqnarray}
\end{subequations}
This is justified if the field envelope $\mathcal{E}$ does not change on length scales comparable to the wavelength $\lambda=2\pi/k_0$ and on time scales comparable to the optical cycle $T_{cyc}=2\pi/\omega_0$.

Strictly speaking, evaluating the third-order nonlinear polarisation vector $\vec{P}^{(3)}$ for optical pulses requires detailed knowledge of the frequency dependence of the nonlinear susceptibility across the relevant spectral range [see Eq.~(\ref{chi3})]. However, this dependence can generally be neglected when the laser frequency lies far from any material resonances.
In this case, we adopt a commonly used approximation for the third-order nonlinearity in gases, expressed as
\begin{subequations}
\label{chi3nlpulse}
\begin{align}
\vec{P}^{(3)}(t) & = \epsilon_0 \sqrt{\frac{\omega_{0}\mu_{0}}{2k_{0}}} \, 2 n_b n_2  I_{\rm eff}(t)\vec{\mathcal{E}}(t) e^{\mathrm{i}(k_{0}z-\omega_0 t)} + \text{c.c.}, \\
 I_{\rm eff}(t)&= \int \! \mathcal{R}(t-t^{\prime}) \left|\mathcal{E}(t^{\prime})\right|^2 dt^{\prime},\\
\mathcal{R}(t) & = (1 - x_{dK}) \delta(t) + x_{dK} \frac{1 + \Omega^2 \tau_K^2}{\Omega \tau_K^2} \, \Theta(t) e^{-t / \tau_K} \sin(\Omega t),
\end{align}
\end{subequations}
where $\delta$ denotes the Dirac delta distribution and $\Theta$ the Heaviside step function.
This intensity-dependent nonlinear polarisation--commonly referred to as the Kerr response--neglects third-harmonic generation, which is strongly suppressed by the large phase mismatch $\Delta k = 3k_0 - k(3\omega_0)$ (for instance, $\Delta k \approx -5$ cm$^{-1}$ for 800~nm in air).
For noble gases, the Kerr response is purely instantaneous ($x_{dK}=0$), corresponding to the electronic contribution $\sim \delta(t)$, which originates from bound electrons. The associated response time, on the order of a few femtoseconds or less, is much shorter than the pulse durations considered here.
In contrast, for air and other molecular gases, a~delayed component proportional to $\exp(-t/\tau_K)$ appears, weighted by $x_{dK}$. This term represents the response that results from the movement of the nuclei, i.e.\ the ro-vibrational Raman contribution, with a characteristic delay time $\tau_K$ and resonance frequency $\Omega$ comparable to the pulse durations of $\sim 100$~fs considered in this study.

As noted above, at high intensities, one must model both the ionisation of gas atoms or molecules and the feedback of the generated charge carriers on the laser field. Depending on the gas and the field intensity, various photoionisation models may be employed by setting appropriate input parameters. Our implementation intrinsically contains a~simple multiphoton ionisation rate or the more general Perelomov–Popov–Terent’ev (PPT) theory \cite{Perelomov_jetpl_23_924}. These ionisation models provide a cycle-averaged, intensity-dependent photoionisation rate $\mathcal{W}_{PI}(I)$ that governs the temporal evolution of the electron density $\varrho_e$.
For consistency, we also include a second ionisation pathway: avalanche (impact) ionisation due to free electrons that are accelerated by the laser field and induce further ionisation. In gases, this mechanism is typically much weaker than that of photoionisation. The full evolution equation for the electron density is therefore
\begin{equation}
\frac{\partial}{\partial t}\varrho_e(\vec{r},t)
= \left[ \mathcal{W}_{PI}(I) + \mathcal{W}_{AI}(I,\varrho_e) \right] \varrho_{nt}(\vec{r},t),
\label{rhoefirst}
\end{equation}
where $\mathcal{W}_{AI}$ denotes the avalanche ionisation rate and $\varrho_{nt}$ the neutral density\footnote{In CUPRAD, recombination can also be included; for conciseness, we omit it here.}. 
At this stage, we consider only weakly ionised plasma: accounting only for single ionisation. This provides the link 
\begin{equation}
\varrho_{nt}(\vec{r},t)+\varrho_e(\vec{r},t)=\varrho^0(\vec{r}),
\label{eq_rho_0}
\end{equation}
where $\varrho^0(\vec{r})$~is the initial time-independent gas-density profile. Therefore, $\varrho_{nt}$~includes the depletion of neutrals by ionisation. We neglect carrier diffusion, which occurs on $\gtrsim\!1$ ps timescales and thus does not affect femtosecond pulses. This means that the $\vec{r}$-dependence is only obtained through the initial density distribution~$\varrho^0$. Within this framework, the electron density $\varrho_e$ is treated as a~slowly varying quantity, analogous to~$\mathcal{E}$.

The oscillatory motion of the free electrons driven by the electric field is described by using the Drude model. We consider a gas of free electrons embedded in a uniform positive ionic background. This background ensures charge neutrality, so that the macroscopic charge density $\varrho$ as introduced in Eq.~(\ref{Maxwell1}) always satisfies
\begin{equation}
\varrho = 0.
\label{eq_rho}
\end{equation}
Since the ion mass is a few thousand times larger than the electron mass, we neglect the motion of ions and, therefore, their contribution to the current density. Under these assumptions, the dynamics of the free-electron current density $\vec{J}_e(\vec{r},t)$ is governed by
\begin{equation}
\frac{\partial }{\partial t}\vec{J}_{e}+\frac{1}{\tau _{0}}\vec{J}_{e}=\frac{q_e^{2}}{m_{e}}\varrho_e \vec{E}. \label{drude}
\end{equation}
Here $q_e$ is for the electron charge, $\tau_{0}$ the electron collision time, and $m_{e}$ the electron mass. The motion of free electrons is mainly governed by the rapid oscillations of the linearly polarised optical field at frequency $\omega_0$. It is therefore convenient to introduce a~complex-envelope function $\vec{\mathcal{J}}_e$,
\begin{equation}
\vec{J}_e(\vec{r},t) = \sqrt{\frac{\omega_{0}\mu_{0}}{2k_{0}}} \vec{\mathcal{J}}_e (x,y,z,t) e^{\cu(k_{0}z-\omega_0t)} + \textrm{c.c.}, \label{sveaj}
\end{equation}
analogous to Equation (\ref{sveart}). We can solve Equation (\ref{drude}) formally as
\begin{equation}
\vec{\mathcal{J}}_{e} = \frac{q_e^{2}}{m_e \omega_0} \left(-\cu\hat{T}+\frac{1}{\tau _{0}\omega _{0}} \right)^{-1} \varrho_e \vec{\mathcal{E}}, \label{jedirect}
\end{equation}
where $\hat{T}=1+\cu \partial_t / \omega_0$. A slowly varying envelope condition similar to (\ref{sveatime}) can be formulated for the product $\varrho_e\vec{\mathcal{J}_e}$ and reads $|\partial_t \varrho_e \vec{\mathcal{E}} |/\omega _{0} \ll |\varrho_e \vec{\mathcal{E}} |$. Thus, together with $1/\tau_0 \omega_0 \ll 1$, it is justified to simplify Equation (\ref{jedirect}) to
\begin{equation}
\vec{\mathcal{J}}_{e} = \frac{q_e^{2}}{m_e \omega_0} \left(\frac{1}{\tau _{0}\omega _{0}}+\cu\hat{T}^{-1} \right) \varrho_e \vec{\mathcal{E}}, \label{je}
\end{equation}
which is much more convenient for the following derivation.

As in the case of the ionisation rates, it turns out that, for consistency, we have to consider a second contribution to the current density: The generation of free carriers by photo-ionisation. Because this term will lead to a loss current in our final equations, we call it $\vec{J}_{IL}$ and
\begin{equation}
\vec{J}_f(\vec{r},t)=\vec{J}_{e}(\vec{r},t)+\vec{J}_{IL}(\vec{r},t). \label{jcomplete}
\end{equation}
The corresponding complex slowly varying envelopes $\vec{\mathcal{J}}$ and $\vec{\mathcal{J}}_{IL}$ are defined as analogues of $\vec{\mathcal{J}}_e$, cf.\ Eq.~(\ref{sveaj}).

Using the principle of energy conservation, self-consistent expressions can be derived for $\mathcal{W}_{AI}$ and $\vec{J}_{IL}$. The temporal evolution of the local energy density $w$ in the medium is determined by
\begin{equation}
\frac{d}{dt}w(\vec{r},t) = \vec{J}_f(\vec{r},t) \cdot \vec{E}(\vec{r},t).
\end{equation}
Consequently, we can compute the energy transferred to the medium by the pulse at the position $\vec{r}$ in a small volume $\Delta V=\lambda_0^3$ by
\begin{align}
& \quad W_{\Delta V}(\vec{r}) = \int_{0}^{\lambda_0} \int_{0}^{\lambda_0} \int_{0}^{\lambda_0} \int_{-\infty}^{\infty} \vec{J}_f(\vec{r}+\vec{r}^{\prime},t)\cdot\vec{E}(\vec{r}+\vec{r}^{\prime},t) dt d^3r^{\prime} \label{fieldloss} \\
& = \lambda_0^3\int \frac{\mu_0 q_e^{2}}{m_e \tau _{0} \omega_0 k_0} \varrho_e(\vec{r},t) \left|\mathcal{E}(\vec{r},t)\right|^2 + \frac{\omega_{0}\mu_{0}}{2k_{0}}\left[\vec{\mathcal{J}}_{IL}(\vec{r},t) \cdot \vec{\mathcal{E}}^*(\vec{r},t)+\textrm{c.c.}\right] dt. \nonumber 
\end{align}
Terms containing $\cu T^{-1}$ vanish upon integration by parts, while rapidly oscillating components proportional to $\exp[\pm \cu2(k_0z-\omega_0t)]$ do not make a net contribution to the integral. The specific choice of $\Delta V$ is arbitrary: its size must be large compared to the characteristic scale of electron displacement, but small relative to the spatial variation of the slowly varying envelopes. 

However, the energy transferred from the optical field to the medium leads to the ionisation of neutral gas atoms or molecules. We assume that all collisional energy losses--represented by the term $\sim1/\tau_0$ in Equation (\ref{drude})--contribute to avalanche ionisation. Consequently, the total energy deposited in the small volume $\Delta V$ can be written as
\begin{equation}
W_{\Delta V}(\vec{r}) = \lambda_0^3U_i\int\left[ \mathcal{W}_{PI}(I) + \mathcal{W}_{AI}(I,\varrho_e) \right] \varrho_{nt}(\vec{r},t) dt, \label{ionizationen}
\end{equation}
where $U_i$ denotes the ionisation energy of atoms or molecules. The term $\propto \mathcal{W}_{PI}$ accounts for the energy consumption associated with photo-ionisation, while the term $\propto \mathcal{W}_{AI}$ represents the contribution of avalanche ionisation to energy consumption. In the slowly varying envelope approximation, the energy transfer is considered local and instantaneous. Since both ionisation processes occur on time scales comparable to or shorter than the optical cycle, and the free-electron motion is confined to nanometre scales, it is justified to omit the time integration when equating Equations (\ref{fieldloss}) and (\ref{ionizationen}). This yields the local energy conservation relation
\begin{equation}
\frac{\mu_0 q_e^{2}}{m_e \tau _{0} \omega_0 k_0} \varrho_e \left|\mathcal{E}\right|^2 + \frac{\omega_{0}\mu_{0}}{k_{0}}\Re\!\left(\vec{\mathcal{J}}_{IL} \cdot \vec{\mathcal{E}}^*\right) = U_i\mathcal{W}_{PI} \varrho_{nt} + U_i\mathcal{W}_{AI} \varrho_{nt} . \label{energyconservation}
\end{equation}

As stated above, both $\mathcal{W}_{AI}$ and $\vec{J}_{IL}$ can be derived in a self-consistent manner. Equation (\ref{energyconservation}) establishes a direct link between these quantities and the known parameters of the system. However, additional considerations are required to ensure their uniqueness. First, we impose that $\vec{J}_{IL}$ only accounts for energy losses, i.e.\ $\Im(\vec{\mathcal{J}}_{IL}\cdot\vec{\mathcal{E}}^*) = 0$. This is primarily a formal constraint, as there is no physical reason to expect that ionisation affects the optical field in any way other than through absorption. Second, we assume that the current density $\vec{J}_{IL}$ arises exclusively from photo-ionisation and that it is parallel to the electric field $\vec{E}$. Under these assumptions, Equation (\ref{energyconservation}) yields
\begin{equation}
\vec{\mathcal{J}}_{IL} = \frac{k_0 }{\omega_0 \mu_0} U_i\frac{\mathcal{W}_{PI}}{\left|\mathcal{E}\right|^{2}}\varrho_{nt}\vec{\mathcal{E}},\label{jmpa}
\end{equation}
and with $\sigma=\mu_0 q_e^2 / m_e \tau _{0} \omega_0 k_0$
\begin{equation}
\mathcal{W}_{AI} = \frac{\sigma}{U_i \varrho_{nt}} \varrho_e \left|\mathcal{E}\right|^2.
\end{equation}

We now justify the neglect of the term $\nabla \cdot \vec{E}$ in the derivation of the wave equation. Under the present assumptions, Equation (\ref{Maxwell1}) reads
\begin{equation}
\nabla\cdot\vec{E}=-\frac{\nabla\cdot\vec{P}^{(3)}}{\epsilon_{0}\epsilon}, \label{diveorig}
\end{equation}
and our goal is to show that $|\nabla\cdot\vec{E}|\ll|\partial_x E_x|$. Although the expression for the nonlinear polarisation $\vec{P}^{(3)}$ is rather complex, its magnitude scales as $\sim 2 \epsilon_0 n_b n_2 I \vec{E}$. For the gases and intensity ranges considered here, the condition $|n_2 I| \ll n_b = n(\omega_0)$ holds. After some algebra and using (\ref{sveaspace}) one can find that $|\nabla\cdot\vec{P}^{(3)}|/\epsilon_{0}\epsilon\ll|\partial_x E_x|$. Hence, $\nabla\cdot\vec{E}=0$ is justified, and we are able to derive the wave equation
\begin{equation}
\Delta \vec{E} -\frac{1}{c^2}\frac{\partial^2}{\partial t^2} \vec{E} = \mu_{0}\left(\frac{\partial^2}{\partial t^2}\vec{P}+\frac{\partial}{\partial t}\vec{J}_f\right).
\end{equation}
With the slowly varying envelopes and the expressions for polarisation [Equations (\ref{polarization}), (\ref{linpol}), and (\ref{chi3nlpulse})] and current density [Equations (\ref{je}), (\ref{jcomplete}), and (\ref{jmpa})] derived above we get
\begin{align}
& \quad \left(\frac{\partial}{\partial z}+\cu k_{0}\right)^2\mathcal{E} + \Delta_{\perp}\mathcal{E} + \int k^2(\omega_0+\overline{\omega}) \widehat{\mathcal{E}}(\vec{r},\overline{\omega}) e^{-\cu\overline{\omega}t}d\overline{\omega} \nonumber\\
& = -\frac{2 k_0^2 n_2}{n_b}\hat{T}^2 \int \mathcal{R}(t-t^{\prime})\left|\mathcal{E}(t^{\prime})\right|^2 dt^{\prime} \mathcal{E} + \frac{k_0^2}{n_b^2 \varrho_c} \varrho_e \mathcal{E}\label{eq:waveeq}\\
& \quad - \cu k_0\hat{T}\left( \sigma \varrho_e \mathcal{E} + U_i \frac{\mathcal{W}_{PI}}{\left|\mathcal{E}\right|^{2}}\varrho_{nt}\mathcal{E} \right),\nonumber
\end{align}
where $\varrho_c = m_e \epsilon_0 \omega_0^2 / \ q_e^2$ is the critical plasma density.

Our objective is to model the evolution of short laser pulses propagating in gaseous media such as noble gases and air. These pulses propagate along the $z$-axis with a well-defined velocity. In the linear regime, they travel at the group velocity $v_g(\omega_0)=1/k^{\prime}$. It is safe to assume, under the present conditions, that the impact of nonlinear effects on $v_g$ is small. Consequently, the derivative of the field envelope $\mathcal{E}$ with respect to $z$ is primarily governed by pulse motion such that
\begin{equation}
\left|\frac{\partial}{\partial z} \mathcal{E} + k^{\prime} \frac{\partial}{\partial t} \mathcal{E}\right| \ll \left|\frac{\partial}{\partial z} \mathcal{E} \right|. \label{pulseass}
\end{equation}
By being transformed into coordinates moving with the pulse, defined by $\tilde{t}=t-k^{\prime}z$ and $\tilde{z}=z$, the dependence of the field envelope on the propagation variable becomes even weaker. Using Eqs.~(\ref{svea_spacetime}) and (\ref{pulseass}) we find
\begin{equation}
\left|\frac{\partial}{\partial \tilde{z}} \mathcal{E}\right| = \left|\partial_z\mathcal{E}+k^{\prime}\partial_t\mathcal{E}\right| \ll \left|\frac{\partial}{\partial z} \mathcal{E}\right| \ll k_0 \left|\mathcal{E}\right|. \label{svearet}
\end{equation}
For simplicity, the tilde notation will be dropped after this transformation. 
Using the Taylor expansion (\ref{nottingham_eq: ksq}), the left-hand side of Eq.~(\ref{eq:waveeq}) becomes
\begin{equation}
\frac{\partial}{\partial z}\left(1+\cu\frac{k^{\prime}}{k_0}\frac{\partial}{\partial t}-\frac{\cu}{2k_0}\frac{\partial}{\partial z}\right)\mathcal{E} -\frac{\cu\Delta_{\perp}\mathcal{E}}{2k_0} +\frac{\cu k^{\prime \prime}}{2}\left(1+\cu\frac{k^{\prime}}{k_0}\frac{\partial}{\partial t}+\frac{\cu}{3}\frac{k^{\prime \prime \prime}}{k^{\prime \prime}}\frac{\partial}{\partial t}\right)\frac{\partial^2\mathcal{E} }{\partial t^2} . \label{eq:lhswaveeq}
\end{equation}
In gaseous media, it is reasonable to assume $k^{\prime}\approx1/c$, $n_b\approx1$, and we can conveniently replace $1+\cu k^{\prime} \partial_t / k_0$ in (\ref{eq:lhswaveeq}) by the operator $\hat{T}$ introduced earlier. Using the relation (\ref{svearet}), we further neglect in (\ref{eq:lhswaveeq}) the second derivative with respect to $z$.
After applying the inverse operator $\hat{T}^{-1}$, the resulting propagation equation reads
\begin{align}
& \quad\frac{\partial}{\partial z}\mathcal{E} - \frac{\cu}{2 k_0}\hat{T}^{-1}\Delta_{\perp}\mathcal{E} + \cu\frac{k^{\prime \prime}}{2}\frac{\partial^2}{\partial t^2}\mathcal{E} - \frac{k^{\prime \prime \prime}}{6}\frac{\partial^3}{\partial t^3}\mathcal{E} \label{timegood}\\
& = \cu k_0 n_2\hat{T} \int \mathcal{R}(t-t^{\prime})\left|\mathcal{E}(t^{\prime})\right|^2 dt^{\prime} \mathcal{E} - \cu \frac{k_0}{2 \varrho_c}\hat{T}^{-1} \varrho_e \mathcal{E} - \frac{\sigma}{2}\varrho_e \mathcal{E} - U_i \frac{\mathcal{W}_{PI}}{2\left|\mathcal{E}\right|^{2}}\varrho_{nt} \mathcal{E},\nonumber
\end{align}
where we have neglected time derivatives of higher order than three.
It is worth noting that Eq.~(\ref{timegood}) goes beyond the standard propagation equation in the slowly varying envelope approximation \cite{Brabec_prl_78_3282}, which reads
\begin{align}
& \quad \frac{\partial}{\partial z}{\cal E} - \cu \frac{1}{2 k_0}\Delta_{\perp} {\cal E} + \cu \frac{k^{\prime \prime}}{2}\frac{\partial^2}{\partial t^2} {\cal E} \label{lasim_pre_1} \\
& = \cu k_0 n_2 \int \mathcal{R}(t-t^{\prime})\left|\mathcal{E}(t^{\prime})\right|^2 dt^{\prime} {\cal E} - \cu \frac{k_0}{2 \varrho_c}\varrho_e {\cal E}  - \frac{\sigma}{2} \varrho_e {\cal E} - U_i \frac{\mathcal{W}_{PI}}{2\left|\mathcal{E}\right|^{2}}\varrho_{nt} {\cal E} \nonumber \\
& = {\cal F}_{NL}(\cal E)\,.\nonumber 
\end{align}
Equation~(\ref{lasim_pre_1}) follows from Eq.~(\ref{timegood}) by neglecting time derivatives of higher order than two and setting $\hat{T}\equiv 1$.

In conclusion, equations \eqref{timegood} and \eqref{lasim_pre_1} are the central equations for our IR-propagation model. Both are implemented in CUPRAD, and users can switch between them using the parameter \texttt{numerics\_operators\_t\_t-1}. As mentioned in the introduction, CUPRAD operates in cylindrical coordinates $(\rho,z,t)$ with $\rho=\sqrt{x^2+y^2}$. Additionally, the rate equation for ionisation~\eqref{rhoefirst} is solved at the same time to provide~$\varrho_e$ and~$\varrho_{nt}$. The model also depends on several material constants, such as the nonlinear refractive index~$n_2$. CUPRAD intrinsically contains tabulated values for noble gases; however, the user can easily provide parameters for other media.

\section{High Harmonic Generation and Propagation}
\label{High Harmonic Generation and Propagation}

In this section, we present the physical model underlying high-harmonic generation (HHG).
The simulation involves two modules : 1)~the computation of the microscopic source terms of the XUV radiation at all points within the macroscopic interaction volume; 2)~the coherent sum of all microscopic emitters to obtain the macroscopic XUV signal.

HHG in gases is generally studied in the non-relativistic regime because too high intensities disrupt the generation mechanism~\cite{Weissenbilder2022}. An accurate description requires a~quantum-mechanical treatment, since the process involves the interaction of an electronic wave packet with the atomic potential. A~suitable framework is provided by the time-dependent Schrödinger equation (TDSE), which describes the dynamics of the electronic wave packet. The emitted fields--generated from all microscopic volumes--are then coherently integrated from the whole macroscopic medium to yield the far-field XUV response.
The details of the respective physical models used in our code are discussed in the following two subsections.

\subsection{Microscopic Response in XUV regime (TDSE)}
\label{sec:High Harmonic Generation and Propagation:Microscopic Response in XUV regime}
We start with the model of the interaction between a~single atom with the external laser field. This calculation is carried out for all spatial points within the macroscopic volume--the macroscopic computational grid is imposed by the solution of the laser propagation, which already ensures the required sampling (the field variations are small from one point to another). Consequently, the microscopic response is evaluated at a~fixed \emph{macroscopic coordinate} that is accounted parametrically in the TDSE, and omitted in this subsection for clarity as only one microscopic system is considered here. The spatial variable introduced here ($\vec{x}_m$) refers exclusively to the \emph{microscopic system}. The complete microscopic dynamics of interest is governed by the time-dependent Schrödinger equation (TDSE)
\begin{align}
	\cu \frac{\partial}{\partial t} \psi(\vec{x}_m,t) = H(t)\psi(\vec{x}_m,t)\,,
\end{align}
which describes the evolution of the atomic wavefunction $\psi(\vec{x}_m, t)$ under the influence of the external laser field. Note that atomic units are used throughout this subsection. A~fully ab initio solution of this problem would involve all electrons in the atom or molecule and hence require solving the TDSE in a~high-dimensional Hilbert space, which becomes computationally infeasible, in particular if we keep in mind that a microscopic solution is required in each macroscopic grid point. To render the problem tractable while preserving the essential physics of the interaction, we must introduce approximations that significantly reduce its dimensionality. Thus, the present model is restricted to linearly polarised laser fields within the dipole approximation and considers a~single active electron without spin. These assumptions make it possible to solve the TDSE in a~one-dimensional geometry represented on a~Cartesian grid. A~future release--which is currently under development--will bring several additional improvements, including exterior complex scaling~\cite{He2007,Orimo2018}, optimised grid representations~\cite{Rescigno2000,McCurdy2004}, more advanced shaping of the atomic potentials~\cite{Sallai2024}, and, ultimately, the incorporation of a three-dimensional TDSE solver within the model.

In general, the incident laser field is characterised by the scalar potential~$\phi(x_m,t)$ and the vector potential~$A(x_m,t)$. The Hamiltonian that describes an electron subjected to both the external laser field and the atomic potential $V_C$ is given by $H=(p+A)^2/2+V_C-\phi$. Both laser-field potentials possess gauge freedom, which leaves the physical electric field--and the magnetic field--invariant provided by $\E= -\frac{\partial }{\partial t} A - \frac{\partial }{\partial x_m} \phi$. Our solver is implemented in the length gauge ($A = 0$ and $\phi(x_m,t) = -E(t)x_m$, consistently with the dipole approximation), which allows us to efficiently directly use the output of CUPRAD in the TDSE. The Hamiltonian reads
\begin{subequations}
	\label{eq:Microscopic Response in XUV regime (TDSE): Hamiltonian (full)}
\begin{align}
	\label{eq:Microscopic Response in XUV regime (TDSE): Hamiltonian (a)}
	H(t) &= \frac{p^2}{2} + V_C + x_m \E(t) \,, \\
	\label{eq:Microscopic Response in XUV regime (TDSE): SC potential}
	V_C  &= -\frac{1}{\sqrt{x_m^2 + a^2}}\,.
\end{align}
\end{subequations}
The potential used in the present model is the so-called soft-core potential \cite{Rae1994} and depends only on the free parameter~$a$ since we impose the asymptoctic potential as $-Z/|x_m|$ when $ |x_m| \rightarrow \infty$ where $Z$ is the ionic charge (here, $Z=1$). The single free parameter $a$ is determined such that the ground-state energy $E_g$ of the field-free Hamiltonian~$H_0=p^2/2+V_C$ matches the desired target value corresponding to a~given atom.

Regarding the choice of gauge, although physical observables are gauge invariant, it is often convenient to perform calculations in a particular gauge depending on the problem at hand. In addition to the length gauge, the velocity gauge ($\phi = 0$ and $E(t) = - \frac{\partial A(t)}{\partial t}$) is commonly used in atomic physics. Transformation between gauges does not pose conceptual difficulty, as it simply corresponds to multiplying the wavefunction by a phase factor, $\psi'(x_m,t)=\mathrm{e}^{\cu \Lambda(x_m,t)}\psi(x_m,t)$, where $\Lambda(x_m,t)$ is the gauge transformation function associated with the field potentials (see Chapter 6.3 of~\cite{Jackson1999}).

Once the time-dependent wavefunction~$\psi(x,t)$ has been computed, the dipole-acceleration source term entering the coupling with the Maxwell's equations is given by
\begin{align}
		\partial{_t}j_m(t)
    &=
		 \braket{\psi(t)| \nabla ( V_C - \phi(x_m,t) )  - \partial_{t} A(t)|\psi(t)} \notag \\
	&=
		 \Braket{\psi(t)| \frac{\partial V_C}{\partial x_m} + E(t) |\psi(t)}  \,,
\label{eq:Microscopic Response in XUV regime (TDSE): dipole acceleration}
\end{align}
where we have used the property $E = -\nabla \phi  - \partial_{t} A$. Note that $j_m(t)$ is related to the charge current and we already use here a~microscopically averaged quantity--provided by the computation of the expected value--, which is exactly required for the coupling with the macroscopic modules. Subtle details of the derivation, including the relation with the macroscopic form of the Maxwell's equations assumed in~\eqref{maxwell} or full 3D-perspective are provided in \ref{app:source terms}.

Taking the wavefunction $\psi(x,t)$, other quantities in addition to~\eqref{eq:Microscopic Response in XUV regime (TDSE): dipole acceleration} can be extracted to gain deeper microscopic insight, such as the photoelectron spectrum~\cite{Catoire2012}, the volumetric and projected ground-state populations, and the energy distribution associated with ionisation~\cite{Vabek2022}. Further analysis tools may include, for instance, the Gabor transform of the source term. Examples of such analyses are discussed in Section~\ref{Interactive Python Interface to CTDSE}, where the source term provides the bridge to the macroscopic model of the XUV field.

\subsection{XUV Propagation (Spectral Approach)}
The propagation of the XUV beam is treated in the linear regime; consequently, the model
neglects any back-reaction of the XUV field on the medium, in contrast to the IR driving field, for which such effects are included (see Section~\ref{Infrared Propagation}).  
Under these assumptions, the XUV~field can be represented as an integral transform of the macroscopic source terms~$\widehat{\p_t \vec{J}}$ integrated over the interaction volume~$V_i$.

Exploiting the cylindrical symmetry of the problem defined by the propagation axis~$z$, the angular part is irrelevant and is omitted in this section. The far-field distribution in a homogeneous generating medium can be expressed as~\cite{Catoire2016}:
\begin{multline}
		\hat{\vec{\E}}(\rho,z,\omega)
	\approx
		- \frac{\mu_0 \e^{\cu k(\omega)z}}{4\pi} \times \\
		\int_{z_{\text{entry}}}^{z_{\text{exit}}}	
		\frac{
			\e^{-\cu k(\omega)z'}
			\e^{\cu \frac{k(\omega)\rho^2}{2(z-z')}}
			}{
			z-z'
			}
		\int_{\Sigma_T}
			\e^{\cu \frac{k(\omega)\left(\rho'\right)^2}{2(z-z')}}	
			\widehat{\left( \pd{\vec{J}}{t}\right)}
			J_0\left(\frac{k(\omega)\rho \rho'}{z-z'}\right) \rho' \dif \rho' \dif z' \,,
\label{eq:XUV prop:Hankel long homogeneous}
\end{multline}
where $(\rho',z')$ denote the cylindrical coordinates within the generating medium, and $(\rho,z)$ defines, analogously, the coordinates of the far-field observation plane. Due to symmetry, the angular integration has been analytic performed -- leading to the radial contributions weighted by the Bessel function of the first kind~$J_0$ --and only the integration in $\rho'$ rests in the transverse part.
The total integration volume is therefore $V_i = \left[z_{\text{entry}}, z_{\text{exit}}\right] \times \Sigma_T$, where $\left[z_{\text{entry}}, z_{\text{exit}}\right]$ defines the longitudinal extent of the medium and $\Sigma_T$ its transverse cross-section.
The precedent microscopic treatment enters the expression through the time Fourier transform of the macroscopic dipole-acceleration $\widehat{\p_t \vec{J}}$ retrieved from~\eqref{eq:Microscopic Response in XUV regime (TDSE): dipole acceleration} by $\widehat{\p_t \vec{J}}(\vec{r}) = \varrho(\vec{r}) \widehat{\p_t \vec{j}}_{\text{m}}(\vec{r})$. The microscopic source is properly weighted by the gas density~$\varrho$ at the macroscopic coordinate $\vec{r}=(\rho,z)$. This coordinate is also reintroduced in~$\widehat{\p_t \vec{j}}_{\text{m}}(\vec{r})$, consistent with the local driving field from which it is computed. Finally, the additional exponential term in the Hankel transform, $\exp \left( \cu k(\omega)(\rho')^2/2(z-z') \right)$, corresponds to the solution of the paraxial wave equation that introduces the correction beyond the standard far-field diffraction approximation. 
The extended propagator accurately describes Gaussian beam propagation and has been successfully applied to model the focussing properties of XUV beams~\cite{Veyrinas2021}.

When accounting for the spatial variation of the gas within the interaction region, the expression~\eqref{eq:XUV prop:Hankel long homogeneous} must be modified accordingly.
This inhomogeneity affects the local wave number $k$, which now explicitly depends on $\rho'$ and $z'$.
Consequently, in the propagator term, the phase factor
 $k(\omega)z'$ is replaced by $k(\rho,z,\omega) = \int_{z'}^{z_{\text{exit}}}k(\rho',z'',\omega) \dif z''$, where $k(\rho,z,\omega)$ represents the spatially dependent dispersion relation incorporating both refractive and absorption effects.
Taking it into account, the far-field electric field, including the influence of the density modulation, is written
\begin{multline}
		\hat{\vec{\E}}(\rho,z,\omega)
	\approx
		- \frac{\mu_0 \e^{\cu k(\omega)z}}{4\pi}
		\int_{z_{\text{entry}}}^{z_{\text{exit}}}	
		\int_{\Sigma_T}
		\frac{
			\e^{-\cu \int_{z'}^{z_{\text{exit}}} k(z'',\rho',\omega) \dif z''}
			\e^{\cu \frac{k(\omega)\rho^2}{2(z-z')}}
			}{
			z-z'
			}
		\times \\
			\e^{\cu \frac{k(\omega)\left(\rho'\right)^2}{2(z-z')}}	
			\widehat{\left( \pd{\vec{J}}{t}\right)}
			J_0\left(\frac{k(\omega)\rho \rho'}{z-z'}\right) \rho' \dif \rho' \dif z' \,.
\label{eq:XUV prop:Hankel long general}
\end{multline}
Further details concerning our specific implementation--such as the normalisation of the signal accounting for absorption through a~complex refractive index--can be found in the manual of the corresponding module.

Building upon the theoretical framework outlined above for the XUV field propagation, we now turn to the practical aspects of the numerical implementation of the respective modules within the code.

\section{A Modular Multiscale Approach (MMA): Implementations}
\label{A Modular Multiscale Approach (MMA)}
\begin{figure}[!ht]
\centering
\includegraphics[width=0.925\textwidth]{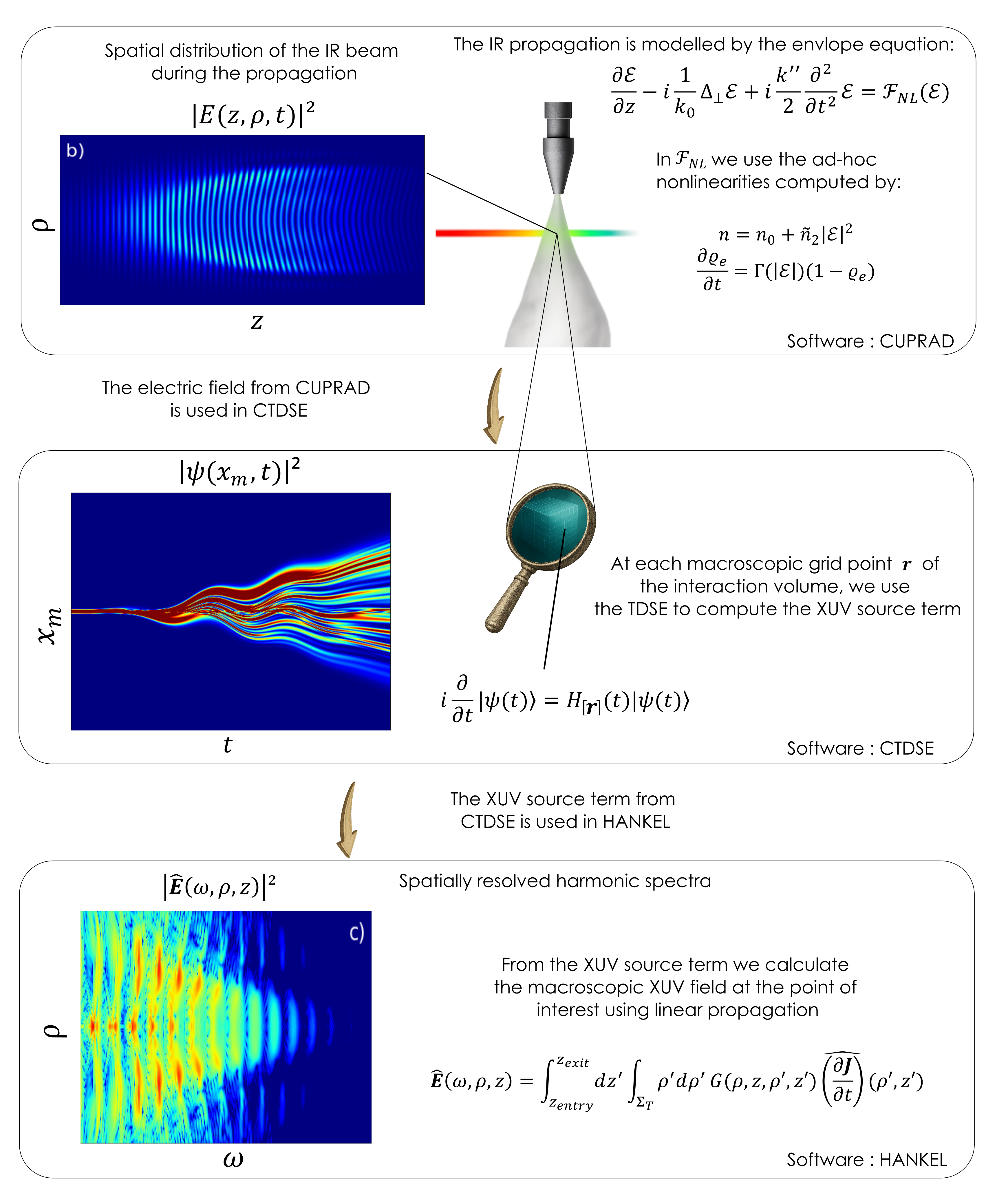}
\caption{The MMA approach unfolds the intrinsically coupled description of the problem outlined in Fig.~\ref{fig:MMA_scheme_intro} into a~pipeline of consecutive tasks. Here we schematically summarise the core ideas of the respective physical models: 1)~the NLSE that governs the propagation of the driving (ionizing) IR field including plasma generation and Kerr effect and provides the laser field in the whole macroscopic volume; 2)~the microscopic TDSE that is solved at each macroscopic point $\vec{r}$ of the medium, describing the atomic microscopic dynamics (spatial variable $x_m$) resulting into the source term in the XUV range; and 3)~the diffraction-integral approach to obtain the macroscopic XUV signal.}
\label{fig:MMA_scheme2}
\end{figure}

The preceding sections have established the physical models that describe all relevant processes.
The next step is to implement it into a~pipeline of numerical solvers that treats consecutively the three main stages of the process, which is schematically illustrated in Fig.~\ref{fig:MMA_scheme2} (cf. the intrinsically coupled general approach in Fig.~\ref{fig:MMA_scheme_intro}).
This is allowed by the spectral decoupling of the driving laser and the secondary radiation. Next, the microscopic feedback loop providing the non-linear propagation is reduced down to scalar response described by the non-linear optics~\cite{Boyd:NLO} and a~simple rate equation to obtain the free electron density, which is computationally tremendous speed-up compared to the numerical solution of~TDSE. The equations within the scheme only represent the central parts of the models, namely the unidirectional equations~\eqref{timegood} and~\eqref{lasim_pre_1} for the IR propagation; 1D-TDSE~\eqref{eq:Microscopic Response in XUV regime (TDSE): Hamiltonian (full)}; and the diffraction integral~\eqref{eq:XUV prop:Hankel long general} for the XUV radiation.
We now proceed to the numerical implementation--the main algorithms and computational methods employed--of the respective modules.
It shall facilitate navigation within the code structure, while further technical details are documented in~\cite{Vabek_Multiscale_modular_approach_2025}.

\clearpage
\subsection{A Parallel Radially Symmetric Unidirectional Solver (CUPRAD)}
Starting with CUPRAD, the simulation of pulsed-beam propagation in gaseous media is technically demanding and typically requires massively parallel computations.
CUPRAD is implemented in Fortran 90 and parallelised for distributed-memory architectures using the Message Passing Interface (MPI) library.
Fast Fourier Transforms (FFTs) are computed using routines from the FFTW3 library.

A split-step scheme is used to solve Equations~(\ref{timegood}) or (\ref{lasim_pre_1}) under the assumption of radial symmetry. The fundamental idea of the split-step (or operator splitting) method is as follows : consider an equation to solve of the form
\begin{equation}
\frac{\partial}{\partial z} \psi = \hat{\mathcal{L}} \psi,
\end{equation}
where $\hat{\mathcal{L}}$ is some operator. Formally speaking, the advance of $\psi(z)$  from $z$ to $z+\Delta z$ is written as
\begin{equation}
\psi(z+\Delta z) = \mbox{e}\,^{\int\!\!\hat{\mathcal{L}}dz}\psi(z),
\label{CUPRAD:general-scheme:sum}
\end{equation}
where $\int$ denotes $\int_z^{z+\Delta z}$. Assuming that $\hat{\mathcal{L}}$ can be decomposed into a sum of $m$ distinct operators,
\begin{equation}
\hat{\mathcal{L}} = \hat{\mathcal{L}}_1 + \hat{\mathcal{L}}_2 + \dots + \hat{\mathcal{L}}_m,
\end{equation}
one may construct a forward split-step scheme
\begin{equation}
\psi(z+\Delta z) \approx \mbox{e}\,^{\int\!\!\hat{\mathcal{L}}_1dz} \mbox{e}\,^{\int\!\!\hat{\mathcal{L}}_2dz} \dots \mbox{e}\,^{\int\!\!\hat{\mathcal{L}}_mdz} \psi(z),
\label{CUPRAD:general-scheme:split}
\end{equation}
and again $\int$ denotes $\int_z^{z+\Delta z}$. 
In general, the operators $\int\!\!\hat{\mathcal{L}}_idz$ do not commute, making~\eqref{CUPRAD:general-scheme:split} an approximation of~\eqref{CUPRAD:general-scheme:sum}, and therefore the order of application of the exponential operators influences the accuracy of the scheme. Comprehensive discussions of higher-order variants of operator splitting can be found in Refs.~\cite{Press_NR_89, Newell_NO_92}.

A key feature of CUPRAD is the adaptative control of the propagation step $\Delta z$ along the longitudinal axis. 
In both Equations~(\ref{timegood}) and~(\ref{lasim_pre_1}), the propagation operator $\hat{\mathcal{L}}$ is decomposed into a linear part, $\hat{\mathcal{L}}_{lin}$, which includes the diffraction and dispersion terms, and a nonlinear part, $\hat{\mathcal{L}}_{nl}$, which contains all remaining contributions.
Since $\hat{\mathcal{L}}_{lin}$ is independent of $z$, a constant step size determined by transverse discretization is generally sufficient.
In contrast, the nonlinear operator $\hat{\mathcal{L}}_{nl}$ exhibits a strong $z$-dependence and thus governs the adaptive step-size control.
To illustrate the principle, consider the simplified example:
\begin{equation}
\frac{\partial}{\partial z} \psi = \cu \frac{\partial^2}{\partial t^2} \psi + \cu \left| \psi \right|^2 \psi.
\end{equation}
Here we have $\hat{\mathcal{L}}_{lin} = \cu \partial_t^2$ and $\hat{\mathcal{L}}_{nl} = \cu \left| \psi \right|^2$. 
Two conditions constrain the choice of $\Delta z$: i- both propgagation steps $\exp(\int\!\!\hat{\mathcal{L}}_{lin}dz)$ and $\exp(\int\!\!\hat{\mathcal{L}}_{nl}dz)$ must remain numerically stable. 
ii- $\Delta z$ must be sufficiently small to preserve the accuracy of the split-step approximation.
The linear step is evaluated numerically exactly in the Fourier space $\exp(\int\!\!\hat{\mathcal{L}}_{lin}dz)=\textrm{FFT}^{-1}\exp(-\cu \omega^2\Delta z)\textrm{FFT}$. 
Thus, there is no restriction on $\Delta z$ from this scheme. 
For the nonlinear step, we approximate the integral $\int\!\!| \psi |^2dz \approx | \psi |^2 \Delta z$, whose relative error scales as $\sim | \psi |^4 \Delta z^2$. 
To control this error, the step size is constrained by $c_2 < | \psi |^2 \Delta z < c_1$, 
where $c_1$ ensures sufficient accuracy, while $c_2<c_1$ promotes a maximally large step-size. 
The split-step method requires that the magnitudes of $\Vert \hat{\mathbb{I}} - \exp(\int\!\!\hat{\mathcal{L}}_{lin}dz) \Vert \sim \Delta z/\Delta t^2$ and $\Vert \hat{\mathbb{I}} - \exp(\int\!\!\hat{\mathcal{L}}_{nl}dz) \Vert \sim | \psi |^2 \Delta z$ remain sufficiently small, where $\Delta t$ denotes the discretisation step of the transverse coordinate  $t$. 
This introduces a~third constant $c_3>\Delta z/\Delta t^2$, which provides an additional upper bound on $\Delta z$ independent of $|\psi|^2$.
Since $\Delta t$ remains constant throughout the simulation, $c_3$ defines a fixed secondary limit on the propagation step.
In practice, the parameter set $c_1 = 0.01$, $c_2 = c_1 / 2.5$, and $c_3 = 1$ has been found to yield stable and accurate results, provided that the transverse discretisation is sufficiently fine.
These conditions are similar to the well-known Courant–Friedrichs–Lewy (CFL) stability criterion~\cite{Courant1928}, which defines the convergence limits for finite-difference schemes by constraining the relationship between spatial and temporal discretisation steps.

CUPRAD solves Equations~(\ref{timegood}) and~(\ref{lasim_pre_1}) in cylindrical coordinates, where the transverse coordinates $x$ and $y$ are transformed as $\rho=\sqrt{x^2+y^2}$ and the angular part is omitted due to the assumed symmetry.
The operator in Equation (\ref{lasim_pre_1}) is decomposed into three parts:
\begin{eqnarray}
\hat{\mathcal{L}}_{lin1} & = & \frac{\cu}{2k_0\rho}\frac{\partial}{\partial \rho}\left(\rho\frac{\partial}{\partial \rho}\right),\hspace*{2.cm}\hat{\mathcal{L}}_{lin2} \,\; = \,\; -\cu k^{\prime \prime} \frac{\partial^2}{\partial t^2}, \nonumber \\
\hat{\mathcal{L}}_{nl} & = & \cu k_0 n_2 \int \mathcal{R}(t-t^{\prime})\left|\mathcal{E}(t^{\prime})\right|^2 dt^{\prime} - \cu \frac{k_0}{2 \varrho_c}\varrho_e - U_i \frac{\mathcal{W}_{PI}}{2\left|\mathcal{E}\right|^{2}}\varrho_{nt}. \nonumber
\end{eqnarray}
The diffraction term, $\hat{\mathcal{L}}_{lin1}$, is treated using a Crank-Nicholson finite-difference scheme (see Ref.~\cite{Press_NR_89}), while the dispersion term $\hat{\mathcal{L}}_{lin2}$ is evaluated in the Fourier space domain.
The non-linear operator, $\hat{\mathcal{L}}_{nl}$, is applied through direct multiplication by $\exp(\int\!\!\hat{\mathcal{L}}_{nl}dz)$, where the integral is approximated as $\int_z^{z+\Delta z}\!\!\hat{\mathcal{L}}_{nl}dz \approx \hat{\mathcal{L}}_{nl}(z)\Delta z$. 
The time integration for the delayed Kerr response and for the electron density [see Equation~(\ref{rhoefirst})] is performed using a simple Euler scheme~\cite{Press_NR_89}, which is sufficient provided that the temporal resolution is fine enough.
Transparent boundary conditions are imposed in the radial coordinate $\rho$~\cite{Hadley1991}.
Since FFT-based propagation requires periodic boundaries in time, fully transparent boundaries in $t$ are not possible; however, the inclusion of approximately $(\sim 16)$ absorbing layers yields satisfactory results, provided that the outgoing field intensity at the edges of the simulation window remains small.
Finally, parallelisation of CUPRAD is implemented via straightforward domain-decomposition, enabling efficient scaling on distributed-memory architectures.

\subsection{1D-TSDE Implementation (CTDSE)}
\label{Implementations: CTDSE}
The numerical solution of the TDSE is implemented in the C programming language.
We conventionally call this implementation~CTDSE (the Computation of Time-Dependent Schrödinger Equation) referring to our specific approach.
It is built around the core routine, which is employed in two distinct applications compiled from the same source.
The first is an MPI-based scheduler that processes the output generated by CUPRAD.
This task is embarrassingly parallel: the scheduler simply distributes macroscopic grid points—each associated with different input field parameters—uniformly across multiple CTDSE processes.
The second application is a dynamically linked C library that provides callable access to the CTDSE routines.
This library is described in detail in Section~\ref{Interactive Python Interface to CTDSE}.

We now outline the principal numerical methods implemented in the core CTDSE routine. 
The wavefunction is represented directly on discrete grids in both space and time.
The spatial and temporal grids are: $[-N_{x}, \dots , 0, \dots , N_{x}] \times \Delta x$~and $[0, 1, \dots, N_{t}] \times \Delta t$. 
For consistency within CTDSE, the temporal grid always begins at $t = 0$.
Consequently, the module internally performs a shift of the input interval $[t_{\mathrm{min}},t_{\mathrm{max}}] \to [0,t_{\mathrm{max}}-t_{\mathrm{min}}]$ to align with this convention. 
Because the temporal discretization in CTDSE is typically finer than that used in CUPRAD, the input interface applies zero-padding in the Fourier domain to interpolate the driving field to the required temporal resolution~$\Delta t$.

Once the computational grids are defined, the wavefunction is represented as $\psi(x_j,t_i) \equiv \psi_j^{(i)}$ and in vectorised form as $\vec{\psi}^{(i)} \equiv (\psi_1^{(i)}, \psi_2^{(i)}, \dots , \psi_{2N_{x,\mathrm{max}}+1}^{(i)})^\intercal$. 
The propagation from $\psi_j^{(i)} \to \psi_j^{(i+1)}$ is performed in three substeps:
\begin{subequations}
\begin{align}
	\vec{\psi}^{(i,1)} &= \left(M_2 - \frac{\cu \Delta t}{2} \left(-\frac{\Delta_2 }{2} + M_2 V_C \right)\right) \vec{\psi}^{(i)}\,, \label{eq:1D-TSDE Implementation (CTDSE):CN - step 4}\\
	\vec{\psi}^{(i,2)} &= \left(M_2 + \frac{\cu \Delta t}{2} \left(-\frac{\Delta_2 }{2} + M_2 V_C \right) \right)^{-1} \vec{\psi}^{(i,1)} \,, \label{eq:1D-TSDE Implementation (CTDSE):CN - step 2}\\
	\psi_j^{(i+1)} = 
	\psi_j^{(i,3)} &= \e^{-\cu \Delta t \E(t_i) x_j} \psi_j^{(i,2)} \,, & \forall j. \label{eq:1D-TSDE Implementation (CTDSE):CN - step 3}
\end{align}
\end{subequations}
The first two substeps implement the Crank-Nicolson scheme for the field-free Hamiltonian using the Numerov-corrected approximation of the Laplacian. 
The matrices involved are : $M_2$ is the Numerov refinement of the Laplacian,~$\Delta_2$ and represents the second-derivative operator, both being tridiagonal (see Section 3.1 of~\cite{Muller1998} for a~detailed discussion). $V_C$~is the potential energy matrix, which is local and therefore diagonal, evaluated on the computational $x$-grid (see~\eqref{eq:Microscopic Response in XUV regime (TDSE): SC potential}). 
All matrix operations thus involve tridiagonal matrices and the inversion in~\eqref{eq:1D-TSDE Implementation (CTDSE):CN - step 2} can be carried out efficiently. 
The interaction with the external electric field is applied in the third substep, Equation~\eqref{eq:1D-TSDE Implementation (CTDSE):CN - step 3}, using a split-operator approach. Since the electric field typically varies on a~much slower time scale than the numerical time step $\Delta t$ (providing $E(t_i)\approx E(t_{i+1})$), the result is in an efficient split-operator propagation scheme with an accuracy of the order $\mathcal{O}(\Delta t^2)$.

All physical observables of interest are computed directly on the spatial $x$-grid.
Most quantities are evaluated within the C implementation itself to ensure high computational efficiency, while the wavefunction~$\psi$ is also made accessible for user-defined post-processing.
The current implementation does not include absorbing boundary conditions.
Consequently, the spatial grid must be chosen sufficiently large to prevent unphysical reflections at the boundaries.
Moreover, the present version of the code does not employ any predictor–corrector propagation scheme, and the next release will provide a version of CTDSE with adaptive step-size control.
A fixed temporal step size is therefore required to perform the numerical Fourier transform of the time-dependent quantities and ensure convergence (see the CFL criterion).
As described in the following sections, the TDSE simulations are executed for multiple sets of input electric fields.
To ensure consistent numerical convergence across all configurations, the time step~$\Delta t$ is chosen according to the most demanding field (i.e., most rapidly varying) in the data set and also an option of grid expansion over time can be activated.

\subsection{A Linear XUV Propagator (HANKEL)}

Once the source terms $\widehat{\partial_t J}$ have been computed at all macroscopic points, they are supplied as input to the diffraction integral, from which the macroscopic spatial profile of the XUV field is calculated.

The calculation of the Hankel transform is implemented in the Python module~\texttt{Hankel\_transform}.
A direct grid integration method is used to evaluate Equation~\eqref{eq:XUV prop:Hankel long general}.
The model can access tabulated scattering factors, requiring only the type of gas and the source of the scattering data as input.

From a~numerical point of view, the $(\rho, \omega)$ far-field plane is partitioned into subarrays that are distributed and processed in parallel using OpenMPI (through Python’s \texttt{multiprocessing} module).
The module provides an extended interface to the core routines, offering flexibility in data handling that benefits from the Pythonic generators: input data can be streamed directly from memory or read dynamically from the disc during execution.
This functionality is particularly useful for the memory management of large datasets, such as those containing dipole-acceleration values across the full macroscopic volume.
More details on the implementation and input/output procedures are provided directly in the open-source code~\cite{Vabek_Multiscale_modular_approach_2025, Vabek2026_zenodo}.

Finally, the \texttt{Hankel\_transform}~module is part of the multiscale model, but it can be used independently as well. It includes i- a long media routine that implements Equation~\eqref{eq:XUV prop:Hankel long general}, and ii- a~lightweight routine for performing standard Hankel transforms applicable to thin-media configurations.

\section{Model Execution and User Interfaces}
\label{User Interfaces}

Here, we present the execution workflow of the model and describe how it enables the use of advanced input parameters to represent more complex physical configurations. 
We then illustrate the methods for analysing and visualising the resulting data. Finally, we outline the HDF5 data structure employed by the code and tools available for post-processing.

The defined data format constitutes the core interface of the entire project and reflects one of its key design philosophies: all data associated with a~given run are packaged together within a~single archive. The final archive therefore contains the complete set of relevant data generated during the run. In particular, all input parameters are preserved, enabling both post-processing analyses and full reproducibility without requiring any additional contextual information. Consequently, even when default parameters are intrinsically defined within the code, they are systematically recorded in the final file. The flexibility of the data format also allows for the inclusion of additional, user-defined data or metadata to support customized applications.

\subsection{Execution pipeline}
\label{User Interfaces: computational pipeline}

The standard operation executes the modules on a supercomputer as a~sequential pipeline of independent jobs. The workflow begins with an initial HDF5 archive that contains all input parameters. Two interfaces are available to generate this initial file: a Jupyter notebook and a text-to-HDF5 converter. The former eases the understanding of the code operation with examples and templates (see Section~\ref{Physics Examples}) shipped with the code. The latter is convenient for preparing large sets of inputs, enabling repeated code executions such as parameter-space scans or similar batch tasks. Finally, the core interface is only the HDF5 archive itself (see Section~\ref{User Interfaces: HDF5}), and this data standard enables easy customisation of the operation for user applications. Once the input file is prepared, the pipeline is launched through a SLURM~\cite{SLURM} job queue.

The input parameters can be broadly classified into two categories: physical and numerical.
The physical parameters define the simulation setup, including the laser pulse characteristics at the entrance and the properties of the medium. 
In most cases, these inputs are scalar values, although more advanced and customisable examples are presented in Section~\ref{User Interfaces: Customised inputs}. 
The second subset of \emph{physical} parameters corresponds to the \emph{material constants}, which typically remain unchanged throughout a series of simulations. 
In our case, these constants are associated with the selected gas species (such as the ionisation potential or the microscopic potential~\eqref{eq:Microscopic Response in XUV regime (TDSE): SC potential}). 
The code provides default values for all material constants, requiring only the specification of the gas type. 
Nevertheless, users can easily modify any of these constants directly in the input file if needed.
The numerical parameters must be explicitly defined in the current version of the code, as the implementation is designed to accommodate a wide range of applications. 
A detailed description of all input parameters is provided in the accompanying documentation. 
Representative test cases are presented in Section~\ref{Physics Examples}.

\subsection{Customised inputs}
\label{User Interfaces: Customised inputs}
Here, we provide a list of advanced input options that enable the modelling of more sophisticated physical configurations:
\begin{itemize}
\item \emph{User-generated Ionisation Rates:} CUPRAD implements by default a~PPT-like ionisation rate (see Appendix~B.1.2 of~\cite{Bergé2007} for implementation details). Alternatively, the code can use externally defined ionisation data by setting the parameter \texttt{CUPRAD/inputs/ionization\_model=ext}. In this mode, a~table of ionisation rates--providing the rate as a function of the electric-field amplitude--must be supplied and stored in the group \texttt{CUPRAD/ionisation\_model}.
\item \emph{Inhomogeneous Gas Densities and Pre-Ionisation:}The model supports the use of arbitrary gas-density profiles, which can be defined either as a~2D map in $(\rho,z)$ coordinates or as a 1D array along any of these axes. This flexibility enables the simulation of HHG in gas jets and other complex geometries~\cite{Strelkov2017, Finke2024}. In a similar manner, a spatial map of pre-ionisation--representing the initial electron density prior to the laser--medium interaction can also be specified. 
\item \emph{User-generated Input Laser Pulses:} The versatility of the code is further enhanced by allowing arbitrary input laser pulse profiles, specified as functions $\E(\rho,t)$. This capability is demonstrated in Example~\ref{Physical examples: HHG with Bessel-Gauss Beams}.
\end{itemize}

\subsection{Post-processing}

Once the computation is completed, the next step consists of analysing and visualising the results. 
The code provides a Python-based data loader for CUPRAD outputs, which instantiates a class containing the data along with several analysis methods. 
The output of all modules is subsequently processed in Python. 
These methods and their applications are presented in Section~\ref{Physics Examples} and are illustrated in the accompanying Jupyter tutorials that are provided with the code~\cite{Vabek2026_zenodo}.

\subsection{HDF5 Data Structure}
\label{User Interfaces: HDF5}
The main communication channel connecting all modules is the shared Hierarchical Data Format~(specifically HDF5~\cite{The_HDF_Group_Hierarchical_Data_Format}). 
All input and output data are stored within a single archive, that is in what follows referred to as \verb|results.h5|. 
Each module reads the required inputs from its designated group in \verb|results.h5| and writes all the corresponding outputs back into the same file.
This architecture is designed to remain transparent to the standard operation described in the previous sections. 
However, since HDF5~files are both machine- and human-readable, the structure can be easily adapted for specific purposes. 
The organisation of the HDF5 archive is defined through the so-called “h5namelists” (see documentation~\cite{Vabek_Multiscale_modular_approach_2025} for details). Modifying the data structure therefore requires changes only to these three lists, which are consistently used across all associated applications.

In this section, we adopt a descriptive approach and focus on:
i- the structure of the input data for the complete model, and
ii- the final data structure produced after all modules have completed their executions.

\subsubsection{Input archive}

Each module group contains a dedicated \verb|inputs| subgroup that stores all input parameters, including precomputed tables such as ionisation rates or density-modulation profiles. 
These elements were discussed in Section~\ref{User Interfaces: Customised inputs}, where we presented two methods to define inputs: a free-form plain-text list of parameters and a Python-based interface. 
Both approaches ultimately serve as wrappers that package the input data into the HDF5 archive. 
By working directly with the HDF5 files, users can easily bypass or customise the input-handling routines provided to suit specific requirements.

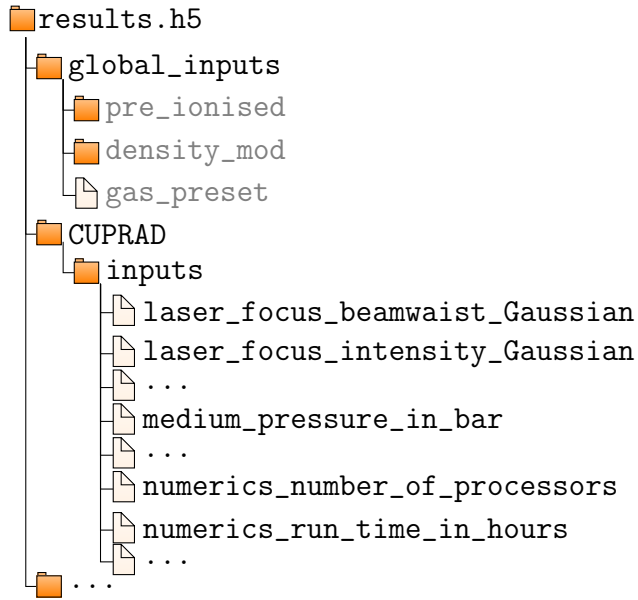
\begin{figure}[ht]
\begin{forest}
  pic dir tree,
  pic root,
  where level=0{}{
    directory,
    inner ysep=-1pt,
    inner xsep=7pt,
  },
[results.h5, inner ysep=-1pt
  [global\_inputs
    	[\textcolor{gray}{pre\_ionised}
    ]
    [\textcolor{gray}{density\_mod}
    ]
    [\textcolor{gray}{gas\_preset}, file, inner xsep=7pt
    	]
  ]
  [CUPRAD
    	[inputs
    	   [laser\_focus\_beamwaist\_Gaussian, file, inner xsep=7pt
    	   ]
    	   [laser\_focus\_intensity\_Gaussian, file, inner xsep=7pt
    	   ]
    	   [..., file, inner xsep=7pt
    	   ]
    	   [medium\_pressure\_in\_bar, file, inner xsep=7pt
    	   ]
    	   [..., file, inner xsep=7pt
    	   ]
    	   [numerics\_number\_of\_processors, file, inner xsep=7pt
    	   ]
    	   [numerics\_run\_time\_in\_hours, file, inner xsep=7pt
    	   ]
    	   [..., file, inner xsep=7pt
    	   ]
    	]
  ]
  [...
  ]
]
\end{forest}
\caption{The HDF5-archive structure showing some of the CUPRAD inputs.}
\label{UI_HDF5_CUPRAD_input_archive}
\end{figure}

We now examine in more detail the inputs required to run CUPRAD, which are stored in the \verb|CUPRAD| and \verb|global_inputs| groups. 
A representative example is shown in Figure~\ref{UI_HDF5_CUPRAD_input_archive}, which illustrates a typical input file used in the simulation described in Section~\ref{Physical examples: HHG in a Gas-Cell}. 
Some inputs are mandatory and must be explicitly provided by the user for each run, whereas others are optional and have predefined hard-coded defaults.During preprocessing, all input data are initialised and stored in the input group. For example, specifying the gas type by storing \verb|Ar| in \verb|global_inputs/gas_preset| automatically triggers the inclusion of hard-coded values for the ionisation potential, dispersion tables, and other related constants. These values are then added to the input group to ensure that the archive always contains a complete, self-consistent record of all simulation parameters. The code also follows an “HDF5 input priority” policy: if a given parameter is present in the input group, its value takes precedence over the corresponding hard-coded default.

\subsubsection{Output archive \& its post-processing}
\begin{figure}[ht]
\begin{forest}
  pic dir tree,
  pic root,
  where level=0{}{
    directory,
    inner ysep=-1pt,
    inner xsep=7pt,
  },
[results.h5, inner ysep=-1pt
  [\textcolor{red}{global\_inputs}
    [\textcolor{pink}{pre\_ionised}
    ]
    [\textcolor{pink}{density\_mod}
    ]
    [\color{pink}{gas\_preset}, file, inner xsep=7pt
    ]
  ]
  [CUPRAD
    	[\textcolor{red}{inputs}
    	]
    	[ionisation\_model
    ]
    [logs
    ]
    [outputs
    ]
    [pre-processed
    ]
  ]
  [CTDSE
    [\textcolor{red}{inputs}
    ]
    [outputs
    ]
  ]
  [Hankel
     [\textcolor{red}{inputs}
     ]
     [outputs
     ]
  ]
  [\textcolor{gray}{warnings}
  ]
]
\end{forest}
\caption{A schematic representation of the archive structure after all modules have completed their computations is shown. The red groups indicate the mandatory inputs required prior to the execution of each module, while the pink groups correspond to optional inputs. The remaining groups contain the output data generated by the respective modules.}
\label{UI_HDF5_output_archive}
\end{figure}
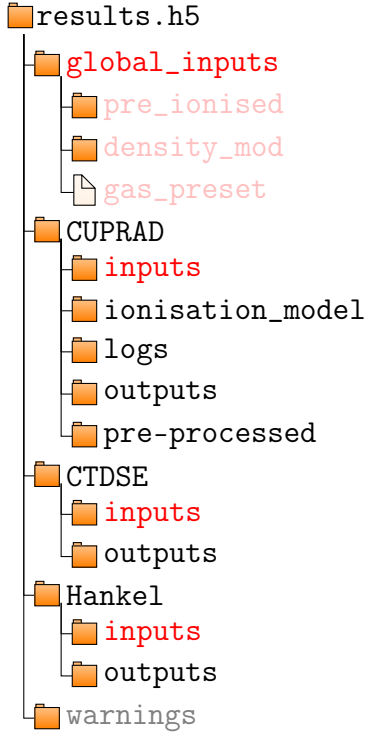

The output archive structure is shown in Fig.~\ref{UI_HDF5_output_archive}. The respective module of the code augments the archive filling the output groups with data. The post-processing of the XUV electric field is straightforward~\eqref{eq:XUV prop:Hankel long general} as it is stored already in the desired form (allowing also options to include cumulative field or separate the influence of the absorption). The post-processing of CUPRAD data is more complex: the electric field and plasma profiles are stored directly in the co-moving frame discussed in Eq.~\eqref{eq:lhswaveeq}. For some applications, one might need to adjust the reference frames or study some derived quantities such as phase or coherence maps (see Section \ref{Physical examples: Absorption-limited HHG with Pre-Ionization in a Gas-Cell}). To make this easily accessible, the CUPRAD module comes with a~Pythonic loader of the output data, it sores them within a~class, which contains directly various methods to process the data (scuh as phase retrieval).

\section{Interactive Python Interface to CTDSE (PyI-CTDE)}
\label{Interactive Python Interface to CTDSE}

The TDSE code can also be accessed independently via a Python interface based on the Ctypes library, which provides access to low-level C functions. 
The underlying computational routines are compiled from the same source as those used in the full-model computational pipeline described in Section~\ref{User Interfaces: computational pipeline}.

Within PyI-CTDSE, users can specify arbitrary input electric fields, offering a convenient and flexible framework for investigating microscopic dynamics. 
In addition to the core solver, the interface includes several analysis tools for processing TDSE outputs, such as the computation of photoelectron spectra, Gabor transforms, time-dependent wavefunctions, time-dependent ionisation probability, etc.

An example of its use is presented in Section~\ref{Physical examples: MIcroscopic insights}. 
Further implementation details, along with the complete list of C-based analysis routines accessible directly from Python, can be found in the documentation of the 1D-TDSE module~\cite{Vabek2026_zenodo}. 
One of the more advanced analyses available is the computation of the time-dependent electron energy density.

\subsection{Analysing the Time-Dependent Electron Energy Density}
\label{Analyzing the Time-Dependent Electron Density}

One of the key tools for gaining detailed microscopic insight is the photoelectron spectrum~\cite{Catoire2012}, which reveals the energy distribution of electrons following their interaction with the laser pulse. 
This spectrum typically exhibits features such as above-threshold ionisation (ATI) peaks, which are crucial for understanding multiphoton and tunnel ionisation processes. 
Extending this analysis to probe the electron dynamics during the interaction requires particular care, due to the gauge freedom inherent in the theoretical description of light–matter coupling. 
To address this, we employ the methodology introduced in~\cite{Vabek2022}, where a detailed treatment of the gauge transformations is provided. 
The resulting quantity is the time-dependent electron energy distribution, defined as
\begin{align}
		\mathcal{P}(E,t) = \lim_{\varepsilon \to 0}\braket{\psi(t)|\hat{R}_{\hat{H}_0}(\varepsilon)|\psi(t)} \,,
		&&
		\hat{R}_{\hat{H}_0}(\varepsilon) = N_{\varepsilon}\varepsilon^2((E-\hat{H}_0)^2 + \varepsilon^2)^{-1}\,,
\end{align}
where $\ket{\psi(t)}$ is the time-dependent wavefunction expressed in the length gauge, $R_{H_0}(\varepsilon)$~denotes the resolvent operator associated with the field-free part of the Hamiltonian~\eqref{eq:Microscopic Response in XUV regime (TDSE): Hamiltonian (a)}, and $N_\varepsilon$ is a normalisation factor specific to the chosen numerical representation. It is often convenient to compute\footnote{Because $H_0$~is the only non-trivial operator in the expression, there is no commutativity issue and we can use $A^2+B^2=(A+\cu B)(A-\cu B)$.}
\begin{align}
		\braket{\psi(t)|\varepsilon^2(E-H_0)^2 + \varepsilon^2)^{-1}|\psi(t)}
	=
		\left\lVert (E-H_0 - \cu \varepsilon)^{-1} \ket{\psi(t)} \right\rVert^2 \,,
\end{align}
where the inversion operation is performed consistently with the numerical scheme used by the solver, as defined in~\eqref{eq:1D-TSDE Implementation (CTDSE):CN - step 2}.

Integrating over energy $\int \mathcal{P}(E,t) \dif E$ provides the time-dependent ionisation probability--or equivalently, the electron density--as directly computed from the TDSE. This quantity can be compared with predictions from alternative models, such as rate-equation approaches.

\section{Physical Examples \& tutorials}
\label{Physics Examples}

In the following, we illustrate the main functionalities of the code suite through representative physical scenarios. 
We begin with standard configurations: HHG in a gas cell (Section~\ref{Physical examples: HHG in a Gas-Cell}) and in a gas jet characterised by a Gaussian density profile (Section~\ref{Physical examples: HHG in a Gas-Jet}). 
We then turn to more advanced cases, including HHG in a pre-ionised gas (Section~\ref{Physical examples: Absorption-limited HHG with Pre-Ionization in a Gas-Cell}), based on a~previous physical application of the code suite~\cite{Finke2022}. 
Subsequently, we demonstrate how user-defined driving pulses can be incorporated (Subsection~\ref{Physical examples: HHG with Bessel-Gauss Beams}). 
Finally, we present a microscopic analysis of the time-dependent electron density (Subsection~\ref{Analyzing the Time-Dependent Electron Density}).

It should be noted that the primary objective of this section is to illustrate the current capabilities of the code suite and the nature of the typical output data it can produce. Each of the examples discussed below is further detailed in a dedicated tutorial provided as a Jupyter notebook shipped with the code.

\subsection{HHG in a Gas-Cell}
\label{Physical examples: HHG in a Gas-Cell}
The default configuration of the code corresponds to a~Gaussian beam and pulse propagating through a gas cell of homogeneous density. 
To illustrate this basic mode of operation, we consider a 30-fs Gaussian pulse (defined as the full width at which the intensity drops to $1/\mathrm{e}^2$ of its peak value) with a waist of $w_0 = 100~\mathrm{\mu m}$. 
The peak intensity is set to $2.44\times10^{14}~\mathrm{W/cm^2}$, which corresponds to a harmonic plateau extending up to the 40th harmonic order according to the classical cut-off law $E_{\text{cut-off}} = I_p + 3.17U_p$~\cite{Corkum1993}. 
Since $E_{\text{cut-off}}$ scales linearly with the intensity through the ponderomotive energy~$U_p$, it provides a convenient reference for specifying the laser intensity, a convention adopted in the following examples. 
The gas cell is 1 ~cm long and filled with argon at a pressure of 25 ~mbar, with the laser focus positioned at the centre of the cell.

\begin{figure}[ht]
\centering
\includegraphics[width=\textwidth]{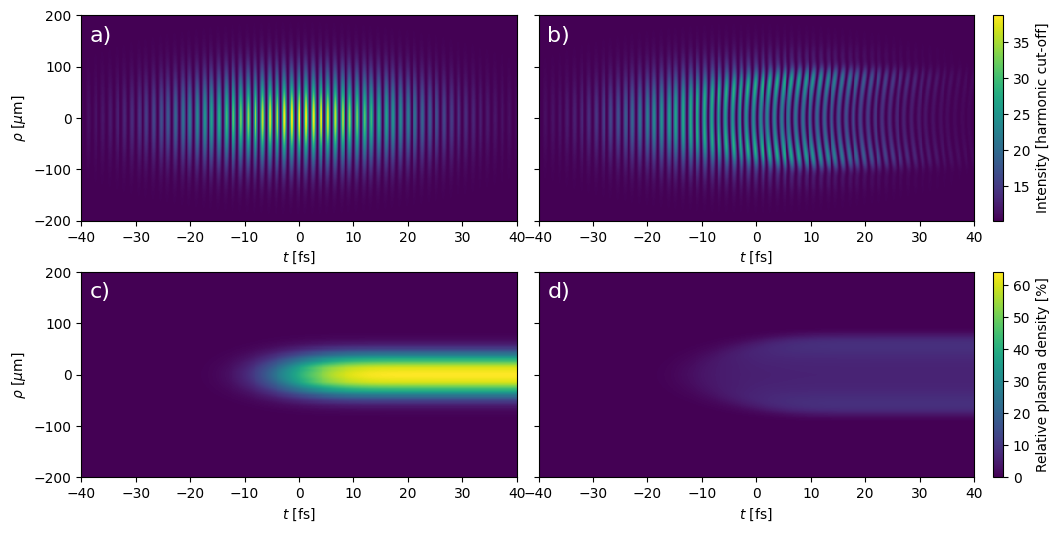}
\caption{The initially Gaussian pulse in time~$t$ and radial coordinate~$\rho$ expressed in the units of harmonic cut-off in panel~a) propagates through 1 cm long and 25 mbar argon gas cell leading to a~distorted pulse~b). The bottom panels~c,d) shows the respective free electron profiles.}
\label{fig:PE_cell_pulse}
\end{figure}

Figure~\ref{fig:PE_cell_pulse} illustrates 1)~the distortion of the laser pulse induced by nonlinear propagation and 2)~the corresponding ionisation dynamics. 
At the exit of the gas cell Fig.~\ref{fig:PE_cell_pulse}(b), the pulse exhibits a reduced peak intensity due to defocusing and a redistribution of energy into the off-axis “wings” of the beam. 
This behaviour results from plasma-induced defocusing: the free electrons generated during propagation, Fig.~\ref{fig:PE_cell_pulse}(c,) locally decrease the refractive index, effectively acting as a diverging lens for the trailing part of the pulse.

\begin{figure}[ht]
\centering
\includegraphics[width=\textwidth]{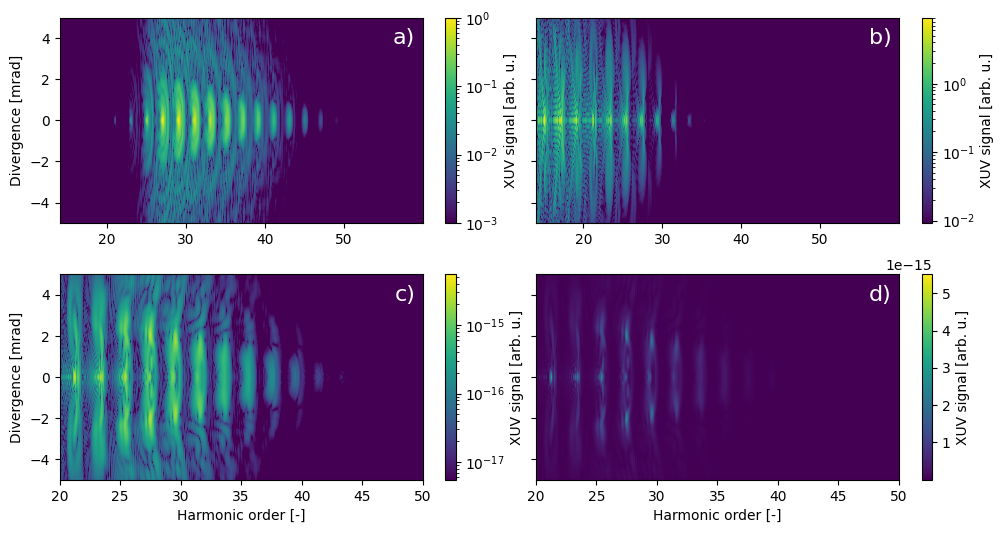}
\caption{Spatially resolved far-field harmonic spectra: a)~and b)~the transforms of the sources in the entry and exit plane respectively, corresponding to the pulse-propagation shown in Fig.~\ref{fig:PE_cell_pulse}; c)~and d)~the harmonic signal integrated along the whole interacting volume in different scales.}
\label{fig:PE_cell_spectra}
\end{figure}

The subsequent stages of the computational pipeline evaluate all dipole responses induced within the macroscopic interaction volume and integrate the resulting XUV emission using a Hankel transform. 
The corresponding harmonic spectra are presented in Fig.~\ref{fig:PE_cell_spectra}. 
For reference, the code can also store the Hankel transforms of both the entry and exit planes. 
Since the input pulse is Gaussian in both space and time, the spatially resolved spectrum obtained by the Hankel transform of the entry plane Fig.~\ref{fig:PE_cell_spectra}(a)--corresponding to the thin medium approximation--exhibits the characteristic ring structures (see also Fig.~8 in~\cite{Catoire2016}). 
The quantity of practical interest--the integrated harmonic spectrum--is displayed at different scales in Figs.~\ref{fig:PE_cell_spectra}(c,d).

In summary, these examples highlight the fundamental operating mode of the code, along with the visualisation of the principal observables typically analysed in HHG simulations.

\subsection{HHG in a Gas-Jet}
\label{Physical examples: HHG in a Gas-Jet}

\begin{figure}[ht]
\centering
\includegraphics[width=\textwidth]{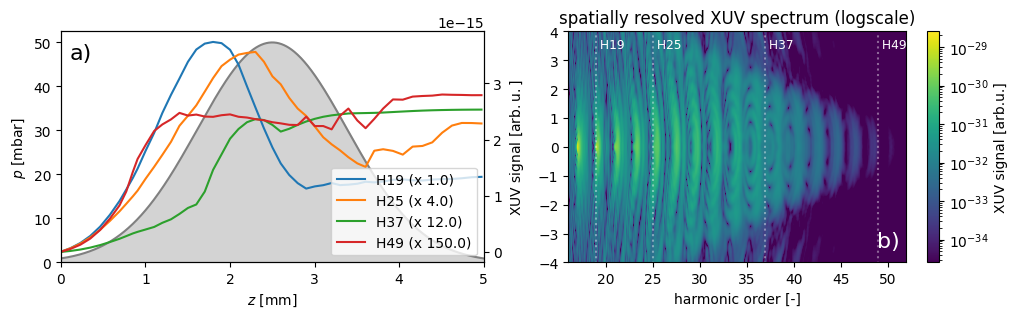}
\caption{The propagation in a~gas jet described in Section~ref{Physical examples: HHG in a Gas-Jet}. a): The pressure profile and the build up of the harmonic signals for the selected harmonics. b) Spatially resolved far-field harmonic spectrum.}
\label{fig:PE_density_profile_XUV_signal}
\end{figure}

This example details the capabilities of the code to simulate pulse propagation in a density-modulated medium. 
A representative case is a gas jet. 
We consider a 15-fs Gaussian pulse with a waist of 100-$\mu$m (parameters defined at focus) centred within a Gaussian pressure profile described by $p(z)=p_0 \exp\left( -(z/a)^2 \right)$ with $a=5/4~\mathrm{mm}$ and $p_0=50~\mathrm{mbar}$.
Figure~\ref{fig:PE_density_profile_XUV_signal}(a) presents the pressure profile together with the cumulative harmonic signals corresponding to the harmonics identified in the spatially resolved far-field spectrum shown in Fig.~\ref{fig:PE_density_profile_XUV_signal}(b).

\subsection{Absorption-limited HHG with Pre-Ionization in a Gas-Cell}
\label{Physical examples: Absorption-limited HHG with Pre-Ionization in a Gas-Cell}
This example is inspired by the full-scale application of the code to the design of an HHG optimisation scheme through homogeneous pre-ionisation of the medium~\cite{Finke2022}. 
In this approach, the presence of free electrons increases the phase velocity of the driving field, allowing it to catch up with the harmonic field and thereby achieve efficient phase matching. 
Rather than reproducing the full physical analysis presented in the reference, here we focus on demonstrating the extended capabilities of the code.

We consider a~45-fs Gaussian pulse with a waist of 100-$\mu$m (parameters defined at focus) centred in a 15-mm-long gas cell filled with krypton at a pressure of $25~\mathrm{mbar}$. 
The medium is initially pre-ionised to $0.9$ times the optimal pre-ionisation level, defined by the condition of equal phase velocities for the fundamental (IR) field and the 17th harmonic~\cite{Finke2022}.

The following analysis demonstrates how more advanced diagnostics can be performed using the complex representation of the pulse envelope, defined as
\begin{align}
	\mathscr{E}_{\text{ce}}(t) = \e^{\cu \omega_0 t} \mathscr{E}_{\text{c}}(t) \,,
	&& 
	 \mathscr{E}_{\text{c}}(t) = \mathscr{F}^{(-1)}\left[ \theta_H(\omega)\mathscr{F}\left[\E(t)\right](\omega) \right](t) \,,
	 \label{PE:absorption-limited:complexified-field}
\end{align}
where $\theta_H$ denotes the Heaviside step function and $\mathscr{F}$ represents the Fourier transform (with spatial coordinates unaffected by the transformation omitted for clarity). 
This complex representation offers an elegant framework for accessing various pulse characteristics. 
The original electric field is retrieved as the real part of the complexified field, $\E(\vec{r},t) = \Re \mathscr{E}_{\text{c}}(\vec{r},t)$. The pulse envelope is given by the modulus of the complex field, $|\mathscr{E}_{\text{c}}(\vec{r},t)| = |\mathscr{E}_{\text{ce}}(\vec{r},t)|$, and the spatio-temporal phase deviation of the field with respect to the monochromatic carrier is obtained as $\Phi(\vec{r},t) = \Arg(\mathscr{E}_{\text{ce}}(\vec{r},t))$.

\begin{figure}[ht]
\centering
\includegraphics[width=\textwidth]{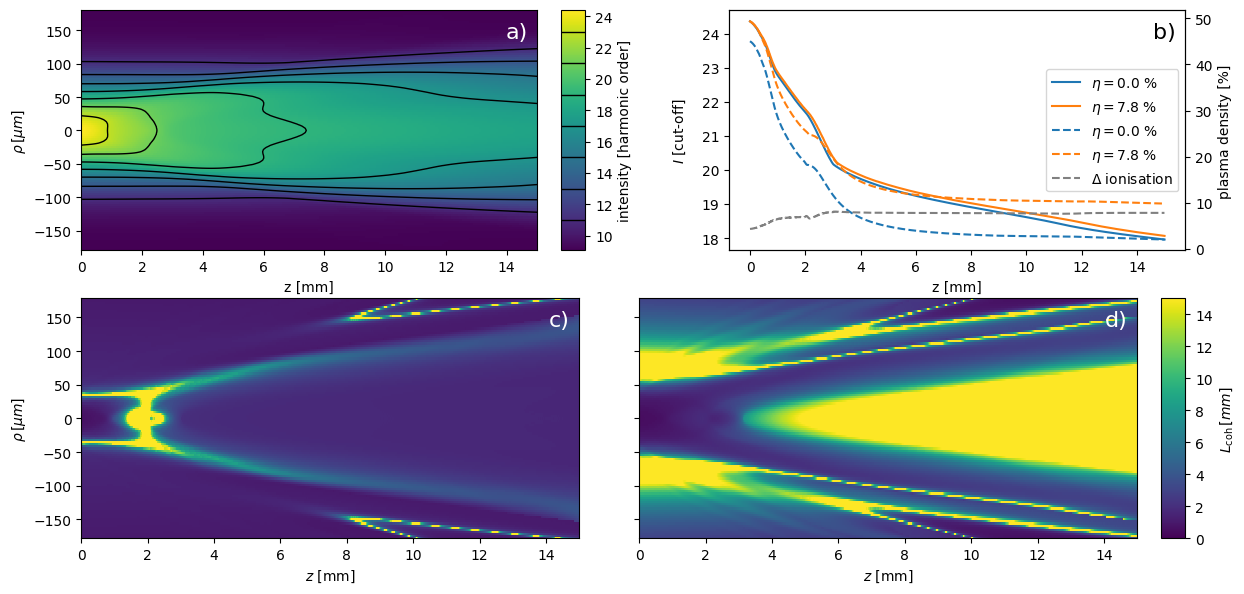}
\caption{a):~The spatial maxima of the intensity profile $\max_t I_{\text{IR}}(z,\rho,t)$ according to the parameters described in Section~\ref{Physical examples: Absorption-limited HHG with Pre-Ionization in a Gas-Cell}. b):~The on-axis ($\rho=0$) intensity profiles with and without the pre-ionisation given by~$\eta$ (full-lines). The $\eta$-value maximise~\eqref{PE:absorption-limited:Lcoh} for an idealised plane-wave propagation. The ionisation degrees for these cases after the passage of the pulse (corresponding dashed lines) and $\Delta$~ionisation as the difference between these two values. The bottom panels~c,d) show the coherence maps~\eqref{PE:absorption-limited:Lcoh} evaluated at times corresponding to the intensity maxima.}
\label{fig:PE_coherence_maps}
\end{figure}

Figure~\ref{fig:PE_coherence_maps}a) presents the spatial peak intensity profile $\max_t |\mathscr{E}_{\text{ce}}(\vec{r},t)|^2$. 
The results reveal intensity clamping within the first third of the medium, followed by a stable, gradual decrease throughout the remaining propagation length. 
This latter region is of particular interest, as it predominantly contributes to the harmonic generation process~\cite{Finke2022}. 
Figure~\ref{fig:PE_coherence_maps}b) further shows that homogeneous pre-ionisation has negligible influence on the propagation dynamics; the additional pre-ionisation merely introduces an almost uniform increase in the electron density within the region of interest.

Figures~\ref{fig:PE_coherence_maps}c,d) present the coherence maps for the 17th~harmonic. 
This metric~\cite{Delfin1999} quantifies the spatial extent of the macroscopic region in which the microscopic emitters interfere constructively. 
It is determined from three main contributions: 1)~the local dephasing of the driving field $\partial_z \Phi$; 2)~the dispersive properties of the harmonic field $\Delta k_{\text{XUV}}=(\omega_{17}/c)(n(\omega_{17})-1)$ derived from the refractive index~$n(\omega)$; and 3)~the dephasing of the microscopic dipoles, which can be approximated as linearly proportional to the longitudinal gradient of the driving-field intensity $\Delta k_{\text{dipole}}\approx -\alpha_{\mathrm{H17}}(I_{\text{IR}}) \partial_z I_{\text{IR}}$. The resulting coherence length is, therefore, expressed as
\begin{align}
		L_{\text{coh, H17}}
	=
		\frac{\pi}{\left| \partial_z \Phi + \Delta k_{\text{XUV}} - \alpha_{\mathrm{H17}}(I_{\text{IR}}) \partial_z I_{\text{IR}} \right|}\,.
\label{PE:absorption-limited:Lcoh}
\end{align}
(See the Supplement of~\cite{Finke2022} for a~detailed derivation.) 

The quantities entering~\eqref{PE:absorption-limited:Lcoh} are obtained as follows: $\Phi$~and~$I_{\text{IR}} \sim |\mathscr{E}_{\text{ce}}|^2$ directly from the CUPRAD outputs; $\Delta k_{\text{XUV}}$ is computed from tabulated dispersion data~\cite{Henke_1993_X-Ray_Interactions_Photoabsorption_Scattering_Transmission_and_Reflection_at_E_50-30000_eV_Z_1-92,NIST_database_of_X-Ray_Form_Factor_Attenuation_and_Scattering_Tables}; and $\alpha_{\mathrm{H17}}$~is evaluated by solving a set of nonlinear equations within the framework of the Strong-Field Approximation~\cite{Smirnova2014}. Hence, the coherence length $L_{\text{coh}}$ provides valuable insight into the high-harmonic generation process without requiring a direct TDSE computation. 
In the present case, it indicates that the optimal phase-matching conditions are achieved in the pre-ionised target Fig.~\ref{fig:PE_coherence_maps}d).

In conclusion, we have presented a representative real-world application of the code, accompanied by in-depth analyses derived from the numerical solution of the CUPRAD module.

\subsection{HHG with Bessel-Gauss Beams}
\label{Physical examples: HHG with Bessel-Gauss Beams}
This example illustrates the use of a custom input field. 
In the current release of the code, such a field must be inserted into the appropriate HDF5 group generated by the CUPRAD pre-processor. 
This group stores the complex electric field~$\mathscr{E}_{\text{c}}(t)$ (see~\eqref{PE:absorption-limited:complexified-field}) defined at the entry plane. 
Here, we replace it with a Bessel–Gauss profile. 
High-harmonic generation is not computed in this section as the purpose is solely to demonstrate the standalone use of the CUPRAD module.

The spatial profile of the input electric field corresponding to the Bessel–Gauss pulse is given by~\cite{Gori1987}:
\begin{align}
		\E(\rho , t)
	=
		\E_0
		\e^{-\left(\frac{\rho}{w_0}\right)^2-\left(\frac{t}{\tau}\right)^2}
		J_0\left(\frac{2r\rho}{w_0}\right)
		\cos\left(\omega_0 t\right)\,,
\label{PE:Bessel-Gauss:focal-field}
\end{align}
where $J_0$~is the Bessel function of the first kind. 
The parameter~$r$ defines the portion of the Bessel contribution to the pulse (the usual Gaussian beam is retrieved for $r=0$).

\begin{figure}[ht]
\centering
\includegraphics[width=\textwidth]{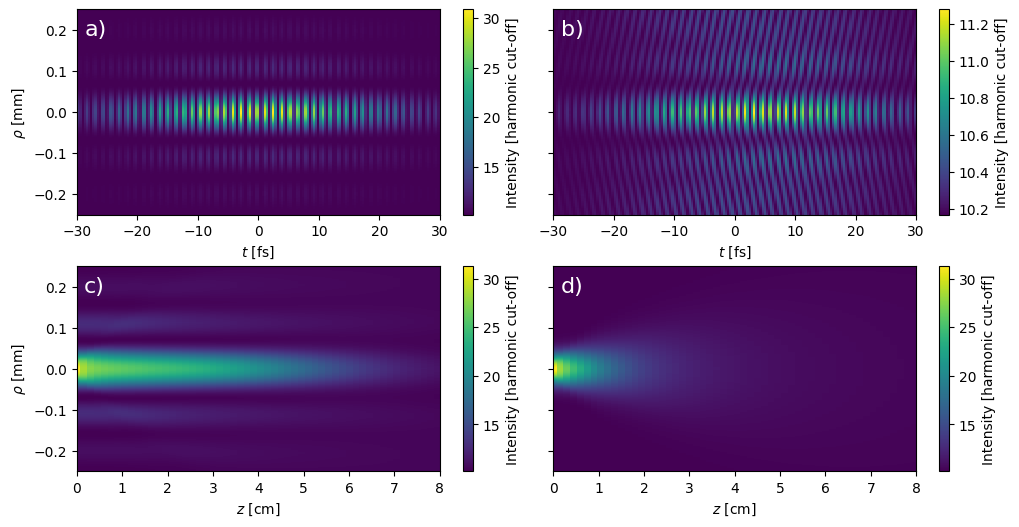}
\caption{The characterisation of the Bessel-Gauss pulse propagation. a):~The initial intensity profile corresponding to the Bassel-Gauss pulse at focus with the parameters described in Section~\ref{Physical examples: HHG with Bessel-Gauss Beams}. b):~The pulse after the propagation through the 8-cm argon cell. c): The peak-intensity profile along the propagation, d):~The purely Gaussian pulse with the same waist as for the Bessel-Gauss pulse.}
\label{fig:PE_Bessel_beam_shapes}
\end{figure}

We consider a~30-fs Bessel-Gauss pulse with a peak intensity of $I_0 \approx 1.7 \times 10^{14}~\mathrm{W/cm^2}$, focused at the entrance of an 8-cm-long gas cell filled with argon at a pressure of 25~mbar. 
The Bessel contribution is set to $r=5$ and the corresponding waist~$w_0$ is adjusted so that the full Bessel-Gauss beam—waist (defined by the $1/e^2$-drop in intensity) is $50~\mathrm{\mu m}$. 
Figure~\ref{fig:PE_Bessel_beam_shapes} illustrates the propagation of this pulse. 
The input profile is shown in Fig.~\ref{fig:PE_Bessel_beam_shapes}a). 
After propagation, the pulse retains its central core, but develops a distinct planar, off-axis, divergent component. Figures~\ref{fig:PE_Bessel_beam_shapes}c),d) compare this behaviour with that of a purely Gaussian beam ($r=0$) having the same input waist, demonstrating that the Bessel-Gauss beam exhibits a significantly extended focal region.

In conclusion, this example highlights the flexibility of the code in handling arbitrary input pulse profiles.

\subsection{Microscopic insights}
\label{Physical examples: MIcroscopic insights}
Finally, we address the microscopic solver, which constitutes both an integral component of the macroscopic computational pipeline and a standalone module accessible directly from Python, as described in Section~\ref{Interactive Python Interface to CTDSE}. 
In the following, we present an example that illustrates the analysis of the 1D-TDSE output.

\begin{figure}[ht]
\centering
\includegraphics[width=\textwidth]{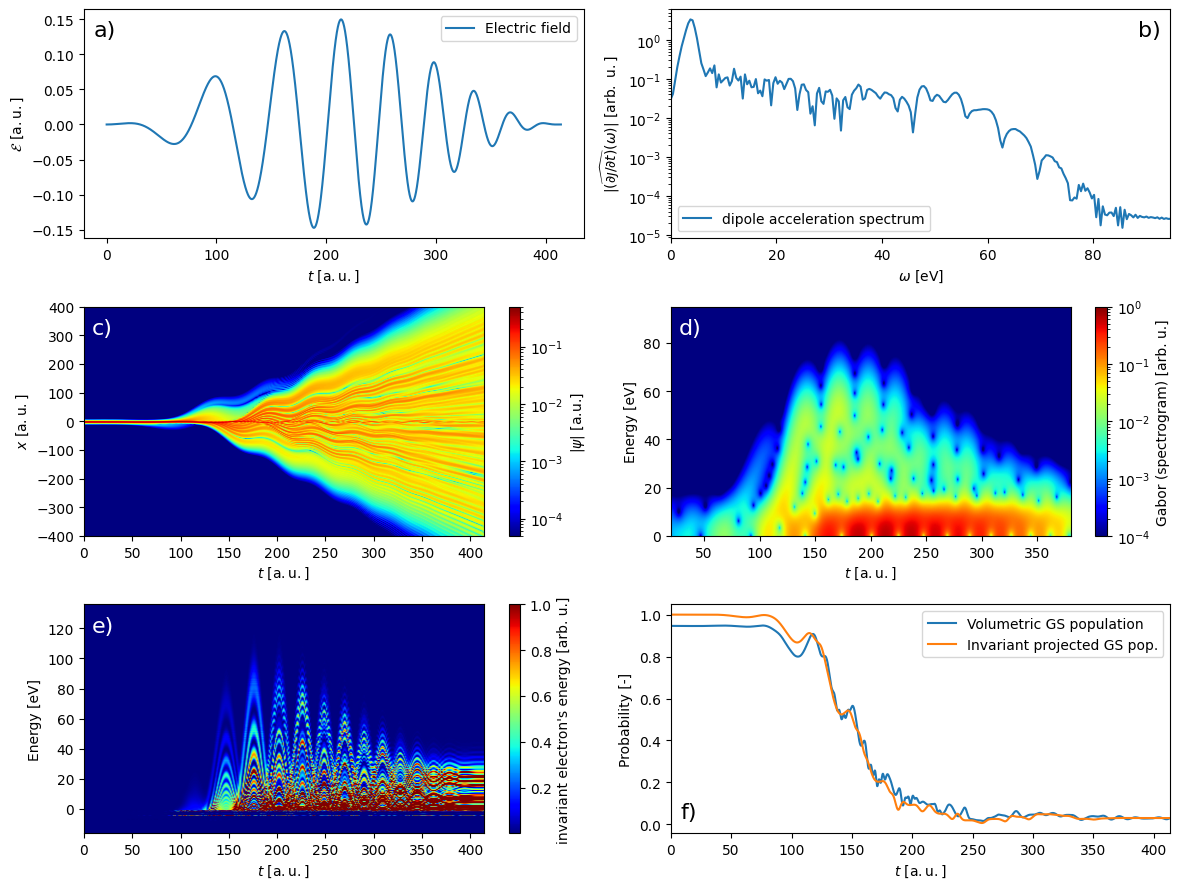}
\caption{The analysis of the 1D-argon atom interaction with the chirped pulse defined in panel~a). The dipole-acceleration~\eqref{eq:Microscopic Response in XUV regime (TDSE): dipole acceleration} spectrum is shown in panel~b). Panel~c) shows the wavefunction modulus $|\psi(x,t)|$, note the various portions of the wavepacked driven by the oscillating field. Panel d)~shows the Gabor transform of $\partial_t j$. Panel~e) shows the time-resolved energetic distribution of the photoelelectron and panel~f) the corresponding populations of the ground state.}
\label{PE:interactive_TDSE_examples}
\end{figure}

We consider the example of a chirped pulse as shown in  Fig.~\ref{PE:interactive_TDSE_examples}a) interacting with an argon atom represented by the soft-Coulomb potential~\eqref{eq:Microscopic Response in XUV regime (TDSE): SC potential}. 
The corresponding dipole acceleration, shown in Fig.~\ref{PE:interactive_TDSE_examples}b), constitutes the source term of the emitted XUV field~\eqref{eq:Microscopic Response in XUV regime (TDSE): dipole acceleration}. 
Further insight into the interaction dynamics is obtained by analysing the wavefunction, the absolute of which is shown in Fig.~\ref{PE:interactive_TDSE_examples}c). 
One such diagnostic is the Gabor transform Fig.~\ref{PE:interactive_TDSE_examples}d), which provides a joint time–frequency representation of the process. 
In simple terms, it reveals when a given frequency--hence the corresponding XUV photon energy--is generated within the driving pulse. 
In the context of the HHG process, the characteristic arc structures indicate the contributions of short trajectories in the rising edge of the pulse, reaching the cut-off region at their apices, and descending along the recombination paths of the long trajectories~\cite{Lynga1999}.
A more advanced analysis is presented in Fig.~\ref{PE:interactive_TDSE_examples}e), showing the photoelectron energy distribution (see~\cite{Vabek2022} for the methodology), along with the time-dependent ground-state population and the volumetric probability of finding the electron in the vicinity of the nucleus, cf.\ Fig.~\ref{PE:interactive_TDSE_examples}f).

In summary, the Python-based TDSE solver and its analysis suite offer a versatile and efficient platform for exploring microscopic dynamics.

\section{Conclusions and perspectives}
\label{Conclusions}

We have introduced a modular code suite designed primarily to model high-harmonic generation (HHG) in noble gases. 
The first public release comprises three main modules:
\begin{enumerate}
	\item CUPRAD--a Fortran-based solver for infrared (IR) pulse propagation in macroscopic media with cylindrical symmetry. 
    It supports linearly polarised pulses and employs MPI parallelisation for large-scale simulations.
	\item CTDSE--a one-dimensional time-dependent Schrödinger equation (1D-TDSE) solver for computing the microscopic response throughout the macroscopic domain. 
    Implemented in C, it includes a Python interface and features an embarrassingly parallel design when integrated into the multiscale framework via MPI.
	\item XUV diffraction--a Python-based module that performs the diffraction integral to coherently sum the microscopic emitters. 
    It uses multiprocessing parallelisation for efficient post-processing.
\end{enumerate}
The suite is complemented by Python pre- and post-processing utilities that simplify the definition of physical setups and the visualisation of simulation results. 
The modules communicate through the HDF5 data format, which ensures efficient interoperability and preserves data accessibility in a human-readable structure.

Despite the simplifying assumptions of this initial release, the suite has already demonstrated its practical applicability to a range of HHG-related problems, producing results in collaboration with experimental teams. 
The project is under active development, with several enhancements and extensions already in progress.
The forthcoming releases will include:
\begin{itemize}
\item an optimised 1D-TDSE solver employing reduced-dimensional potentials,
\item a 3D-TDSE module for arbitrarily polarised pulses,
\item SFA-like microscopic models (both using saddle point in momentum or full saddle points for analytical input fields),
\item additional numerical and modelling features,
\item 3D IR pulse propagation with arbitrary polarisation,
\item XUV propagation to arbitrary distances--which will be solved using an excact propagator--, and
\item hollow core fibre propagation.
\end{itemize}
These developments will be accompanied by expanded documentation, additional tutorials, and practical tools already in use in the developers’ institutions--ensuring that feature prioritisation aligns with real experimental and theoretical needs. 
The broad range of tools aims to provide an optimal balance between physical accuracy and computational efficiency.

A central goal of the project is to guarantee both scientific reliability and user accessibility of the open-source package:
\begin{itemize}
\item \emph{User perspective:} The suite is distributed with Jupyter-notebook tutorials, a Docker container for local deployment, and installation guides for HPC clusters. 
The package has been successfully tested on multiple architectures with both Slurm and PBS schedulers. All documentation is included with the first public release available in the project’s Git repository~\cite{Vabek_Multiscale_modular_approach_2025, Vabek2026_zenodo}.
\item \emph{Developer perspective:} A well-defined software architecture--with explicit module interfaces and reproducible computational pipelines--provides a solid foundation for further community-driven contributions in computational HHG and related fields.
\end{itemize}
In summary, this project establishes a collaborative framework that aims to connect theorists, numerical physicists, and experimentalists. 
The code suite is intended as a practical and extensible platform for both fundamental studies and applied designs of XUV beamlines~(e.g.~\cite{Hort2019}), allowing users to model, optimise and interpret HHG experiments within a unified computational environment.


\FloatBarrier
\appendix

\section{Microscopic source terms}
\label{app:source terms}
The goal of this appendix is to provide further physical context on the assumptions used for modelling the non-linear source term for CUPRAD and computed from TDSE via Eq.~\eqref{eq:Microscopic Response in XUV regime (TDSE): dipole acceleration}. 
The main difference comes already from the initial assumptions in the derivation of CUPRAD, starting from the macroscopic version of Maxwell's equation~\eqref{maxwell}, which already contains the microscopically averaged source terms~\cite{Russakoff1970} assumed in Eqs.~\eqref{Maxwell1} and~\eqref{Maxwell4}.

The main differences arise from the distinct spectral and intensity ranges relevant to IR pulse propagation on the one hand, and that of HHG in the XUV range on the other.
We begin with IR propagation, whose spectral extent lies below the ionisation threshold. 
In this regime, the dominant contributions stem from processes involving i- bound electrons, which drive the nonlinear polarisation terms ~$\vec{P}^{(i)}$ in Eqs.~\eqref{linpol} and~\eqref{chi3nlpulse}, and ii- free electrons, which govern ionisation and recombination processes described in Eq.~\eqref{jcomplete}.
In principle, capturing these effects would require a full quantum-mechanical description of the interaction, including a 3D solution of the TDSE and the multielectron dynamics. (Note that the dimensionality of the~3D-solution of $N$-particle TDSE is~$3N$, which makes an exact ab initio solution beyond the computational power available today.)
In practice, however, these contributions are accurately and efficiently modelled using standard linear and nonlinear optical response theories, which provide reliable and extensively validated tools for this regime.

The situation is markedly different for HHG. This process originates from the well-known three-step model~\cite{Corkum1993}, in which a single active electron is sufficient and essential to capture the underlying physics. 
Remarkably, the dominant features of HHG are already well reproduced within a reduced one-dimensional geometry when the driving field is linearly polarised.
In this context, the exact sub-cycle structure of the electric field plays a crucial role, and this temporal structure is naturally and accurately incorporated in the 1D-TDSE framework.

Let us now examine more closely the microscopic source terms appearing in Eq.~\eqref{eq:Microscopic Response in XUV regime (TDSE): dipole acceleration} and their connexion to the macroscopic Maxwell equations in~\eqref{maxwell}.
To do so, we explicitly invoke Ehrenfest’s theorem, which can be written as:
\begin{align}
		\frac{\partial}{\partial t} \braket{\hat{O}(t)} = \Braket{\frac{\partial}{\partial t}\hat{O}(t)} + i\braket{[\hat{O}(t),H(t)]}
\label{eq:Ehrenfest}
\end{align}
in order to relate the different physical quantities of interest and to connect those that can be directly computed.

We begin with the dipole moment. Classically, it is given by $q\vec{r}(t)$.
Its quantum analogue is $q\braket{\psi(t)|\vec{r}|\psi(t)}=-\braket{\psi(t)|\vec{r}|\psi(t)}$ where we used $q=-1$ for the electron. 
We also recall that this quantity is gauge invariant.
 Applying Ehrenfest’s theorem successively, we obtain expressions for the microscopic charge current and for its time derivative, namely:
\begin{subequations}
\begin{align}
		\vec{j}_m(t) =& -\braket{\psi(t) |\vec{p} + \vec{A}(t) |\psi(t)}\,,
\label{eq:Current_TDSE} \\
		\partial_t\vec{j}_m(t) =& \braket{\psi(t)|\vec{\nabla} \left( V_c -\phi(t) \right) - \partial_t \vec{A}(t)  |\psi(t)} \notag\\ 
		=& \braket{\psi(t)|\vec{\nabla} V_c  + \vec{E}(t)  |\psi(t)}\,.
\label{eq:Current_der_TDSE}
\end{align}
\end{subequations}
All these quantities are gauge invariant, and we have used the definition of the electric field provided by $\vec{E}(t) = - \partial_t \vec{A}(t)-\vec{\nabla} \phi(t)$.
In our 1D version of the code, the derivative of the current is provided by:
\begin{align}
		\partial_t j_m(t) = \int \psi^*(x,t) \psi(x,t) \left( \partial_x V_c(x) +E(t) \right) \dif x\,.
\label{eq:Current_der_TDSE1D}
\end{align}
The connexion between this microscopic current and its macroscopic counterpart, the total current $\vec{J}=\vec{J}_{f}+\partial_t\vec{P}$ appearing in Eq.~\eqref{Maxwell4}, is based on the assumption of a dilute medium such that all atoms can be treated as independent emitters.
Under this assumption, the time derivative of the total macroscopic current can be written as:
\begin{align}
		\partial_t \vec{J}(\vec{r},t) = \partial_t\vec{J}_f(\vec{r},t) + \partial_t^2\vec{P}(\vec{r},t) = \frac{q_e^2}{m_e} \varrho^0(\vec{r}) \partial_t\vec{j}_{m[\vec{r}]}(t),
\label{eq:Current_der_TDSE1D_rho}
\end{align}
where $\varrho^0$ denotes the initial gas density, see Eq.~\eqref{eq_rho_0}. 
In the current version of the code, the gas density is time-independent, while its spatial dependence is explicitly implemented (see Section~\ref{Physical examples: HHG in a Gas-Jet}). 
The microscopic source term $\vec{j}_m$ depends parametrically on the macroscopic position $\vec{r}$ through the local value of the driving field used in the TDSE.
Note that the quantum-mechanical computation of the expected value provides the averaging over the microscopic systems assumed in~\cite{Russakoff1970}. 
Furthermore, it also naturally contains the dynamics of the ionisation process.

Having defined the total macroscopic current $\vec{J}$, can we still infer~$\vec{P}$ and~$\vec{J}_f=\vec{J}_e+\vec{J}_{IL}$ that appear in Eq.~\eqref{Maxwell4} and in the macroscopic pulse propagator from the microscopic current computed from the TDSE? 
Following the theory presented in~\cite{Vabek2022}, and thus working in the length gauge, the wavefunction can be decomposed into bound and continuum components as 
$$\ket{\psi(t)} = \ket{b} + \ket{c}.$$
With this decomposition, the current can be written as 
\begin{equation*}
\begin{split}
\partial_t\vec{j}_m(t) & = \braket{c|\vec{\nabla} V_c  + \vec{E}(t)  |c} + \braket{b|\vec{\nabla} V_c + \vec{E}(t)  |b} \\
&\quad  + \braket{b|\vec{\nabla} V_c  + \vec{E}(t)  |c} + \braket{c|\vec{\nabla} V_c  + \vec{E}(t)  |b}.
\end{split}
\end{equation*}
This expression allows one to identify the contributions from the different transitions.
In particular, the free-electron contribution is given by $$\partial_t\vec{j}_e= \braket{c|\vec{\nabla} V_c  + \vec{E}(t)  |c}$$ and is naturally connected to the macroscopic free-current term $\vec{J}_e = \frac{q_e^2}{m_e} \rho^0\vec{j}_e$, cf.\ Eq.~\eqref{eq:Current_der_TDSE1D_rho}. 
This current contains ~$\braket{c|c}\vec{E}(t) = p_{\rm ion}(t)\vec{E}(t)$, where $p_{\rm ion}(t)$ is the ionisation probability, corresponding to classical Brunel radiation~\cite{David2025, Brunel1990}. The remaining part given by  $\braket{c|\vec{\nabla}V_c|c}$ is associated with potential-induced radiation and can be neglected~\cite{David2025} so that $$\partial_t \vec{J}_e \approx \frac{q_e^2}{m_e}\rho_e\vec{E}(t).$$
Here, compared to Eq.~\eqref{drude} in Section~\ref{Infrared Propagation}, the damping term due to collisions is not included because the TDSE considers an isolated atomic system without coupling to an environment. The ionisation dynamics can also be extracted independently using the resolvent method, as described in Sect.~\ref{Analyzing the Time-Dependent Electron Density}.

The remaining contributions to the total current involve the bound electrons and correspond to $\vec{P}$ and $\vec{J}_{IL}$. In particular, neglecting continuum contributions to $\vec{P}$ we can link it to the bound-bound term as $$\partial_t^2\vec{P}\approx\frac{q_e^2}{m_e} \rho^0 \braket{b|\vec{\nabla} V_c  + \vec{E}(t)|b}.$$ 
The mixed terms corresponding to bound-continuum and continuum-bound transitions, $\braket{b|\vec{\nabla} V_c  + \vec{E}(t)  |c} + \braket{c|\vec{\nabla} V_c  + \vec{E}(t)  |b}$, describe, among other effects like HHG, the losses of the IR laser field due to ionisation. In the IR propagator, the ad hoc expression for $\vec{J}_{IL}$ accounts for these losses, cf.\ Eq.~\eqref{jmpa} in Sect.~\ref{Infrared Propagation}. We expect that evaluating the mixed terms in the infrared frequency range reproduces the ionisation loss current $\vec{J}_{IL}$ in the leading order, but we leave the verification of this claim for future work.

In the present implementation of the code, the microscopic source term used for the coupling to Maxwell's equations is Eq.~\eqref{eq:Current_der_TDSE}.
Since the Fourier transform of the source term is used for the coupling to the Maxwell's Equations, one often writes the formal relations $\widehat{\braket{\psi(t)|\vec{r} |\psi(t)}}(\omega)=i\omega\widehat{\vec{j}_m}(\omega)=-\omega^2\widehat{\partial_t\vec{j}_m}(\omega)$. However, these identities must be used with care when performing a Fast Fourier Transform (FFT), because then periodic boundary conditions in time are imposed. 
In such cases, the dipole, current, and acceleration forms are no longer strictly equivalent in the frequency domain.

\section{Acknowledgements and author contributions:}
\emph{J.V.}~is the main designer of interfaces between the modules, the architect of the data-structure. He participated in the development of all the modules and adapted them for specific tasks. He is the main author of the Hankel module. He designed i)~a~part of the main CTDSE routines and their parallelisation, and ii)~computational pipelines and tutorials.
\emph{T.N.} ~refactored the c-code for the CTDSE solver and developed the Pythonic interface of CTDSE; he worked on the deployment of the code, the installation tools, and the containerisation of the codebase.
\emph{S.S.}~is the main author of CUPRAD, including core routines, parallelisation, and physical model.
\emph{F.C.}~is the main author of CTDSE and the physical model of the Hankel module.
All authors contributed to the writing of the manuscript, documentation, and reviewed the work on the whole project, including physical models and code reviews.

\emph{J.V.}~acknowledges 1)~all the collaborators who contributed to the development, in particular Jakub Jelínek, who implemented a part of the hdf5 routines in Fortran, and Edwin Chacón-Golcher, who supported and optimised the deployment of the code on ELI cluster; 2)~the support from the project Advanced Research Using High Intensity Laser Produced Photons and Particles (CZ.02.1.01/0.0/0.0/\-16\_019/0000789) from the European Regional Development Fund (ADONIS) and support from GACR Grant No. 25-17853S.

\emph{S.S.} acknowledges all colleagues who have contributed to the development of the CUPRAD code, in particular R. Nuter, who developed the ionisation module, and L. Berg\'e for extensive testing, many suggestions for improvements, and invaluable discussions on the physical model. 

All authors acknowledge various providers of computer time during the development: 1) the SUNRISE cluster at the ELI-Beamlines facility. 2)~This work was supported by the Ministry of Education, Youth and Sports of the Czech Republic through e-INFRA CZ (ID:90254). 3)~The computer time for this study was provided by the MCIA (Mésocentre de Calcul Intensif Aquitain) and 4)~the Grand Equipement National de Calcul Intensif (GENCI project A0XX0507594).
Finally, we acknowledge the feedback of all our collaborators on the applications of the code~\cite{Veyrinas2021} and~\cite{Finke2022}.

 \bibliographystyle{elsarticle-num} 
 \bibliography{cas-refs}





\end{document}